\documentclass[superscriptaddress, aps, prl,twocolumn]{revtex4-1}

\usepackage{graphicx}
\usepackage{amsmath}
\usepackage{gensymb}
\usepackage{siunitx}

\usepackage{makecell}
\usepackage{dcolumn}
\usepackage{multibib}

\newcommand{\beq}{\begin{equation}}
\newcommand{\eeq}{\end{equation}}

\newcommand{\beqa}{\begin{eqnarray}}
\newcommand{\eeqa}{\end{eqnarray}}

\usepackage[usenames, dvipsnames]{color}
\usepackage{multirow}

\graphicspath{ {Figures/} }

\usepackage[mathcal]{eucal}

\begin{document}

\title{Linear hyperfine tuning of donor spins in silicon using hydrostatic strain}

\author{J. Mansir}
\affiliation{London Centre for Nanotechnology, UCL, 17-19 Gordon St, London, WC1H 0AH, United Kingdom}

\author{P. Conti}
\affiliation{London Centre for Nanotechnology, UCL, 17-19 Gordon St, London, WC1H 0AH, United Kingdom}

\author{Z. Zeng}
\affiliation{Universit\'e Grenoble Alpes, CEA, INAC-MEM, L\_Sim, F-38000 Grenoble, France}

\author{J.J. Pla}
\affiliation{School of Elec. Engineering \& Telecomm., University of New South Wales, Sydney, NSW 2052, Australia}

\author{P. Bertet}
\affiliation{Quantronics Group, SPEC, CEA, CNRS, Universit\'e Paris-Saclay, CEA-Saclay, 91191 Gif-sur-Yvette, France}

\author{M.W. Swift}
\affiliation{Materials Department, University of California, Santa Barbara, California 93106-5050, USA}

\author{C. G. Van de Walle}
\affiliation{Materials Department, University of California, Santa Barbara, California 93106-5050, USA}

\author{M.L.W. Thewalt}
\affiliation{Department of Physics, Simon Fraser University, Burnaby, British Columbia, Canada V5A 1S6}

\author{B. Sklenard}
\affiliation{Universit\'e Grenoble Alpes, CEA, INAC-MEM, L\_Sim, F-38000 Grenoble, France}

\author{Y.M. Niquet}
\affiliation{Universit\'e Grenoble Alpes, CEA, INAC-MEM, L\_Sim, F-38000 Grenoble, France}

\author{J.J.L. Morton}
\affiliation{London Centre for Nanotechnology, UCL, 17-19 Gordon St, London, WC1H 0AH, United Kingdom}
\affiliation{Dept of Electronic and Electrical Engineering, UCL, London WC1E 7JE, UK}

\begin{abstract}

	We experimentally study the coupling of Group V donor spins in silicon to mechanical strain, and measure strain-induced frequency shifts which are linear in strain, in contrast to the quadratic dependence predicted by the valley repopulation model (VRM), and therefore orders of magnitude greater than that predicted by the VRM for small strains $|\varepsilon| < 10^{-5}$.
	Through both tight-binding and first principles calculations we find that these shifts arise from a linear tuning of the donor hyperfine interaction term by the hydrostatic component of strain and achieve semi-quantitative agreement with the experimental values. Our results provide a framework for making quantitative predictions of donor spins in silicon nanostructures, such as those being used to develop silicon-based quantum processors and memories. The  strong spin-strain coupling we measure (up to 150~GHz per strain, for Bi-donors in Si), offers a method for donor spin tuning --- shifting Bi donor electron spins by over a linewidth with a hydrostatic strain of order $10^{-6}$ --- as well as opportunities for coupling to mechanical resonators.
\end{abstract}

\maketitle

Donors in silicon present an attractive spin qubit platform, offering amongst the longest coherence times in the solid-state~\cite{Tyryshkin11, Saeedi13} and single-qubit control with fault-tolerant fidelity~\cite{Pla13,Muhonen15}. 
As with the conventinal semiconductor industry, the majority of efforts in donor-based spin qubits are focused on $^{31}$P donors~\cite{Kane98,Obrien01, Tyryshkin03, Stegner06, Morton08, Simmons11, Fuechsle12, Buch13}. The heavier group V donors $^{75}$As, $^{121}$Sb, and $^{209}$Bi have recently received substantial interest~\cite{Franke15,Franke16,Wolfowicz14,Singh16,Morley10,Mortemousque14,Saeedi15}, offering larger nuclear spins (up to $I=9/2$ for $^{209}$Bi) and correspondingly richer Hilbert spaces that enable up to four logical qubits to be represented in a single dopant atom. Furthermore, ``atomic clock transitions'' have been identified in $^{209}$Bi where spin resonance transition frequencies become first-order insensitive to magnetic field noise, resulting in coherence times of up to 3 seconds in $^{28}$Si~\cite{Wolfowicz13}.

The exploitation of donor spins in silicon as qubits typically requires their incorporation into nano- and micro-electronic devices. This has been used to demonstrate single-shot readout of a single $^{31}$P donor spin using a tunnel-coupled silicon single-electron transistor (SET)~\cite{Morello10,Pla12}, and to create hybrid devices in which donor spins are coupled to superconducting resonators~\cite{Bienfait16,Eichler17,Zollitsch15} to develop interfaces between microwave photons and  solid-state spins. 
In both cases, the use of metal-oxide-semiconductor (MOS) nanostructures~\cite{Angus07}, or patterned superconducting films on silicon~\cite{Bruno15} involves a combination of materials with coefficients of thermal expansion that differ by up to an order of magnitude \cite{Lyon77,Swenson83,Roberts82,NIST91,NixMacnair41}. The presence of strain in the silicon environment around the donor spin is therefore pervasive when studying such nanodevices at cryogenic temperatures. Furthermore, factors such as optimising spin-resonator coupling or spin-readout speed motivate the placement of donors close to features such as SETs~\cite{Morello09} or resonator inductor wires~\cite{1608.07346} where strain is maximal.

Strain modifies the band structure of silicon \cite{BardeenShockley50,HerringVogt57}, as has been shown, for example, to contribute to the confinement of single electrons in silicon under nanoscale aluminium gates \cite{Thorbeck15,Veldhorst15}. 
The donor electron wavefunction is also modified by strain: following the valley repopulation model (VRM) developed by Wilson \& Feher \cite{WilsonFeher61} within the framework of effective mass theory, an applied uniaxial strain lifts the degeneracy of the six silicon valleys leading to a mixture of the donor ground state, $1s(A_1)$, with the first excited state, $1s(E)$. In this excited state, the hyperfine coupling between the donor electron and nuclear spin is zero, therefore the VRM predicts a quadratic reduction in $A$ as a function of uniaxial strain, as well as a strain-induced anisotropic contribution to the electron g-factor.
Strain-induced perturbations of the donor hyperfine coupling have been observed for P-donor spins in $^{28}$Si epilayers, grown on SiGe to yield built in strains of order 10$^{-3}$~\cite{Huebl06} --- piezoelectric contacts on such material have been used to modulate this built-in strain to shift the electron spin resonance frequency by up to $\sim400$~kHz~\cite{Dreher11}.

In this Letter, we report the observation of a strain-induced shift in the hyperfine coupling of group V donors in silicon which is linear (rather than quadratic), and therefore orders of magnitude greater than that predicted by the valley repopulation model of Wilson and Feher~\cite{WilsonFeher61} for small strains ($|\varepsilon|\lesssim 10^{-5}$). We present experimental studies showing strain-tuning of the coherent evolution of each of the group V donor spins ($^{31}$P, $^{75}$As, $^{121}$Sb, and $^{209}$Bi), extracting the strain-induced shifts of the hyperfine coupling and electron spin g-factor for each, and corroborate the results with a combination of both tight binding and density functional theory calculations which reveal the crucial role of hydrostatic strain in this novel mechanism~\cite{PBS57_main}. In addition to providing essential insights for the use of donor spins in nano- and micron-scale quantum devices, our results provide a method for linear tuning of the donor hyperfine interaction with coupling strengths of up to 150~GHz/strain for $^{209}$Bi donor spins.

\begin{figure}[t]
	\centering
	\includegraphics[width=90mm]{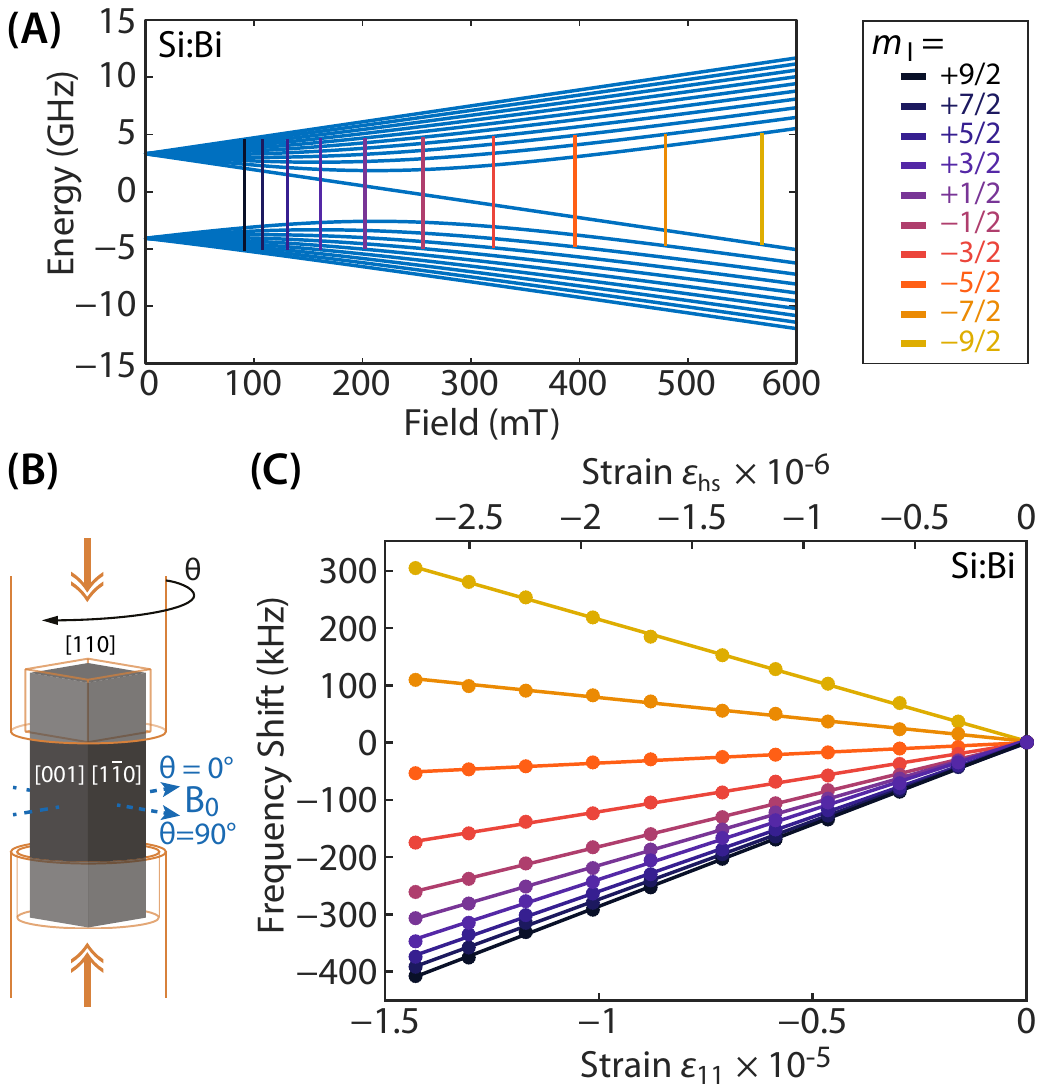}
	\caption{\textbf{(A)} Energy levels of Si:Bi spin eigenstates as a function of magnetic field, with the ten allowed ESR transitions at 9.7 GHz highlighted and  labelled according to their high-field nuclear spin projection $m_I$. \textbf{(B)} Schematic of experimental setup. The silicon single crystal sample is mounted between two engineered plastic rods with the ability to apply compressive stress. $\theta$, the angle of the applied magnetic field to the $[001]$ direction, can be varied by rotating the sample. 
	\textbf{(C)} Observed linear frequency shifts for each of the ten allowed ESR transitions shown in panel (A), as a function of strain $\varepsilon_{11}$, with $\theta = 30\degree$.
	\label{fig:figone}}
\end{figure}

The spin Hamiltonian for a group V donor in the presence of an external magnetic field $\mathbf{B}=B_0\mathbf{z}$ is:
\beq
\mathcal{\hat{H}} =\big(g_e\mu_B\mathbf{\hat{S}_z} - g_n\mu_N \mathbf{\hat{I}_z}\big)B_0 + A\mathbf{\hat{S}}\cdot\mathbf{\hat{I}} 
\eeq
where $g_e$ and $g_n$ are, respectively, the electronic and nuclear g-factors, $\mu_B$ and $\mu_N$ are the Bohr and nuclear magnetons, $\mathbf{\hat{S}}$ and $\mathbf{\hat{I}}$ are the electronic and nuclear spin operators. The Fermi contact hyperfine interaction strength, $A = 1.4754$~GHz in Si:Bi, can be expressed as:
\beq
A = \cfrac{8\pi}{3} g_0\mu_B g_n\mu_N |\psi(0)|^2 
\eeq
where $\psi(0)$ represents the amplitude of the electronic wavefunction at the nucleus and $g_0=2.0023$ is the free electron g-factor.%
The eigenstates of this Hamiltonian describe a Hilbert space of dimension $(2S+1)(2I+1)$, with $S=1/2$ and $I$ determined by the nuclear spin species, illustrated for the case of $^{209}$Bi:Si in Fig.~\ref{fig:figone}(A). Transitions between these eigenstates obeying the selection rule ($\Delta m_{S},\Delta m_{I}) =(\pm1,0)$ in the high field limit can be driven and detected using pulsed electron spin resonance (ESR) \cite{Schweiger05_main}.

\begin{figure}[t]
	\includegraphics[width=87mm]{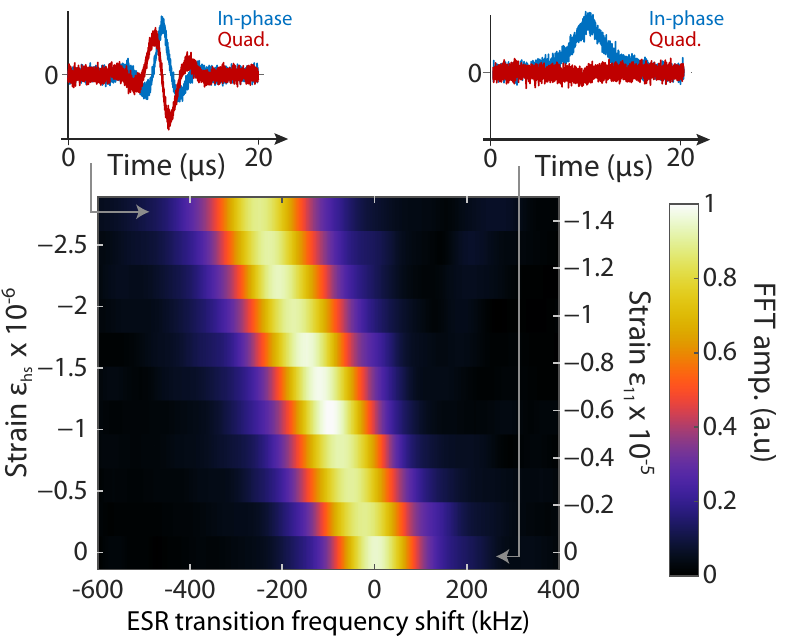}
	\caption{Electron spin echo signals in the frequency domain measured in $^{28}$Si:Bi as a function of compressive strain (shown in terms of the uniaxial strain $\varepsilon_{11}$ and hydrostatic strain $\varepsilon_{\rm hs}$)  arising from the applied stress in our experiment. 
	Time-domain echoes for zero strain and $\varepsilon_{11}=1.4 \times 10^{-5}$ are shown as insets. Data shown is from the $m_I = -1/2$ transition with $\theta = 45\degree$, taken at $T= 8$~K. \label{fig:frequency_domain}}
\end{figure}


We use samples of isotopically enriched $^{28}$Si doped with Bi, Sb, As and P (see Ref~\cite{suppmat} for more details), mounted with crystal orientation shown in Fig.~\ref{fig:figone}(B).
The sample is situated inside a dielectric microwave ESR resonator in an Oxford Instruments CF935 liquid helium flow cryostat, and is held between two PEEK rods whose end faces have been CNC-milled to match the profile of the sample allowing it to be rotated with respect to the magnetic field.
Using calibrated masses, a uniaxial stress is applied to the sample perpendicular to the [110] face, via the upper rod which extends outside the cryostat.
The resulting strain tensor can be derived from the generalised form of Hooke's law for anisotropic materials and the compliance matrix for silicon~\cite{Zhang14}: in the ([110],[1$\bar{1}$0],[001]) coordinate system per kg of applied mass, $\varepsilon_{11} = \num{-1.45e-5}$/kg, $\varepsilon_{22} = \num{9.02e-7}$/kg, $\varepsilon_{33} = \num{5.24e-6}$/kg, and $\varepsilon_{i\neq j} = 0$. While the VRM predicts frequency shifts only from uniaxial strain, we shall see that the new mechanism presented here arises from hydrostatic strain $\varepsilon_{\rm hs} = (\varepsilon_{11} + \varepsilon_{22} + \varepsilon_{33})/3$. In our setup, we estimate a strain per unit mass $\varepsilon_{\rm hs}= \num{-2.78e-6}$/kg.

We use a home-built pulsed ESR spectrometer~\cite{Ross17} at 9.7 GHz to apply a Hahn echo sequence $\pi/2 \rightarrow \tau \rightarrow \pi \rightarrow \tau \rightarrow echo$~\cite{Hahn50}, with $\tau = 15~\mu$s and a $\pi$ pulse duration of 130~ns. 
The time-domain Hahn echo signals (top of Fig.~\ref{fig:frequency_domain}) obtained while systematically increasing the applied strain are then Fourier transformed to yield the strain-induced shifts in spin resonance frequency~\cite{suppmat}.

\begin{figure}
	
	\includegraphics[width=87mm]{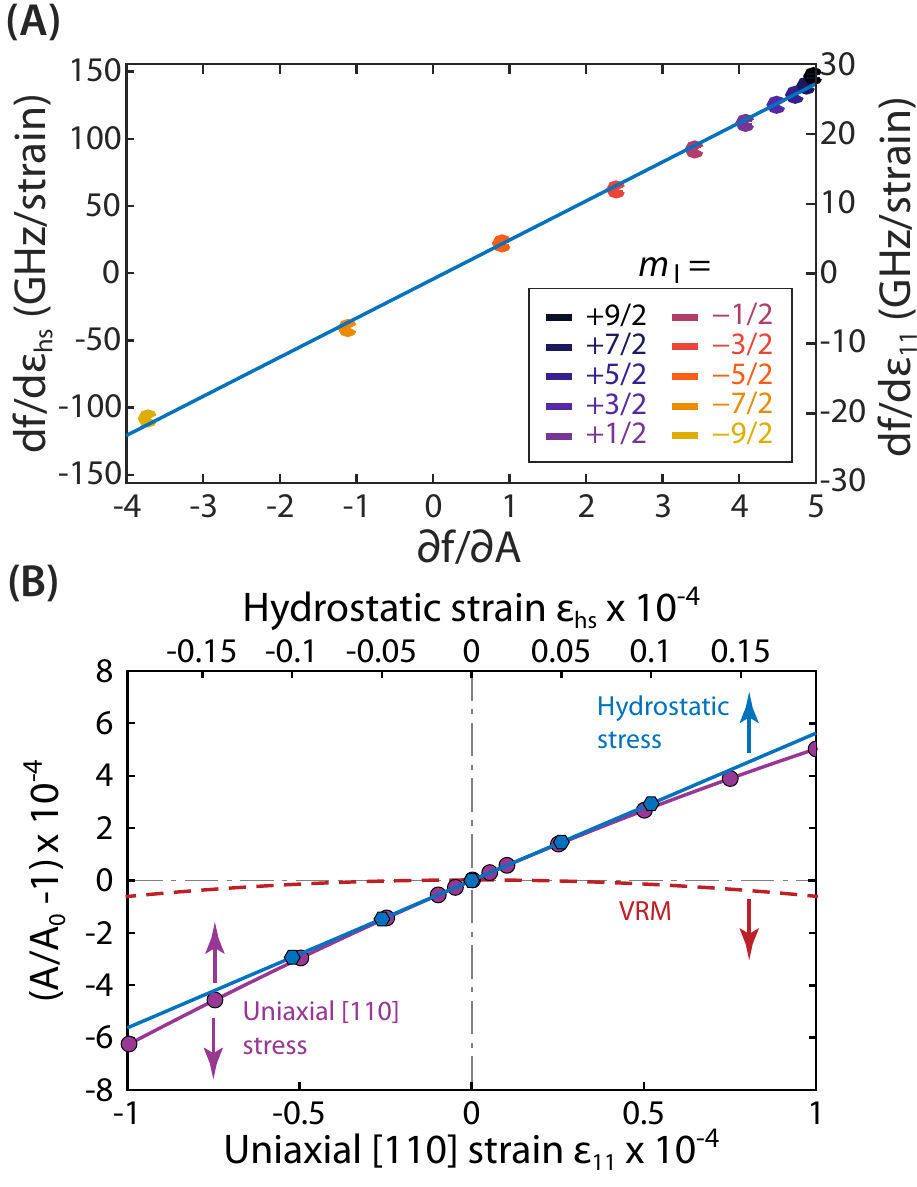}
	\caption{
		\textbf{(A)} The gradient of the strain-induced frequency shifts (d$f/$d$\epsilon$) for each of the ten Si:Bi ESR transitions are shown as a function of the first-order sensitivity of each transition to the hyperfine coupling ($\partial f/\partial A$). The linear relationship confirms the observed strain-induced shifts in Si:Bi result from tuning the hyperfine coupling, $A$, with a gradient $\partial A/\partial \epsilon_{11} = 5.4\pm0.3$ GHz or, equivalently, $\partial A/\partial \epsilon_{\rm hs} = 28.2\pm1.6$ GHz.  
		\textbf{(B)} Calculations showing the relative change in hyperfine coupling strength $A/A_0$ as a function of strain, comparing tight-binding (TB) and  valley repopulation (VRM) models. To mimic the experiment, we show TB calculations of $A/A_0$ under uniaxial stress along $[110]$ (purple curve and circles), which produces a hydrostatic component of strain ($\varepsilon_{\rm hs}$, top axis) in addition to a uniaxial component along $[110]$ ($\varepsilon_{\rm 11}$, bottom axis). This behaviour can be understood by comparing with TB calculations for pure hydrostatic stress (blue curve and hexagons), plotted on the same axis of $\varepsilon_{\rm hs}$, as well as calculations from the VRM (red dotted curve), plotted on the same axis of $\varepsilon_{\rm 11}$. The arrows indicate the relevant axes for each trace. \label{fig:dfda_fits}}
\end{figure}

First, we observe in Fig.~\ref{fig:frequency_domain} that the Bi donor ESR transition can be shifted by more than a linewidth (in $^{28}$Si) for strains of order 10$^{-5}$ (uniaxial) or 10$^{-6}$ (hydrostatic).
We fit the frequency-domain echo signals to a Voigt profile, and then plot the centre-frequency shifts as a function of strain, for each of the ten allowed ESR transitions (see Fig.~\ref{fig:figone}(C)). Strikingly, the ESR frequency of each transition shows a linear dependence on strain, rather than the expected quadratic dependence.
In Fig. \ref{fig:dfda_fits}, we plot the experimentally determined $\partial f/\partial \epsilon_{11}$ for each transition against the first-order sensitivity of each transition frequency to the isotropic hyperfine coupling $\partial f/\partial A$. Remarkably, all 10 points fall on a single line, demonstrating that the dominant effect we observe in Si:Bi is a strain-induced shift in the isotropic hyperfine coupling which is linear in strain, and equivalent to  $\partial A/\partial \varepsilon_{11}$ = $5.4\pm0.3$ GHz or $\partial A/\partial \varepsilon_{\rm hs}$ = $28.2\pm1.6$ GHz.

Multivalley effective mass theory (EMT) has been successful in describing many aspects of the donor electron wavefunction \cite{KohnLuttinger55,LuttingerKohn55, Pantelides74}, including close agreement between theory and experimental measurements of the Stark effect \cite{Pica14} and predictions of exchange coupling between neighbouring donors \cite{Pica14_2}. Within this framework, the wavefunction is expanded in terms of Bloch functions concentrated around the six degenerate [100] conduction band minima (valleys) such that $\psi = \sum_{\mu=1}^6\alpha_\mu F_\mu\phi_\mu$ where  $\mu$ indexes over the valleys in the basis $[+x,-x,+y,-y,+z,-z]$, $F_\mu$ is a hydrogen-like envelope function, and $\phi_\mu$ is the valley Bloch function. The donor impurity potential breaks the symmetry of the crystal and induces a coupling between the valleys, leading to a valley-orbit splitting of the $1s$-like donor state into three sub-levels. The ground state is singly degenerate with $A_1$ symmetry and has $\alpha_\mu = 1/\sqrt{6}$, while one of the excited states is doubly degenerate with $E$ symmetry and has $\alpha_{E_1} = 1/2[1,1,-1,-1,0,0]$ and $\alpha_{E_2} = 1/2[1,1,0,0,-1,-1]$. The valley repopulation model assumes that uniaxial strain applied along a valley axis results in the corresponding pair of valley energies being decreased or increased for compressive or tensile strain, respectively~\cite{WilsonFeher61}. 
This modification of the valley energies results in a redistribution of the amplitude of each valley contributing to the ground state, which can be represented under strain as an admixture of the $1s(A_1)$ and $1s(E)$, resulting in a quadratic reduction of $A$ as a function of uniaxial strain. At our maximum applied strain of $\varepsilon_{11} = -1.45 \times 10^{-5}$, the VRM predicts a reduction in $A$ of 1.9~kHz, while we measure a reduction in $A$ of 78 kHz --- this discrepancy is even more pronounced for smaller strains. Therefore, in addition to predicting a different functional form of the dependence of $A$ against strain, the VRM predicts shifts which are approximately two orders of magnitude smaller than what we measure in this strain regime, implying that another physical mechanism must dominate the changes to the structure of the donor electron wavefunction we observe.


In order to understand these trends, we have computed the bound states of bismuth impurities in silicon using the $sp^3d^5s^*$ tight-binding (TB) model of Ref.~\onlinecite{Niquet09}. This model reproduces the variations of the band structure of bulk silicon under arbitrary strains in the whole first Brillouin zone. The impurity is described by a Coulomb tail and by a correction of the orbital energies of the bismuth atom (similar to a central cell correction in the effective mass approximation) \cite{Usman15}. The TB ratio $A/A_0$ between the strained ($A$) and unstrained ($A_0$) hyperfine interaction strengths is plotted in Fig. \ref{fig:dfda_fits}(B) under uniaxial stress along $[110]$, as a function of the resulting uniaxial $[110]$ and hydrostatic strains. Surprisingly, and in agreement with the experiments performed here, $A/A_0$ behaves linearly with small strain, and this trend can be assigned to the effects of the hydrostatic stress. Although not predicted by the VRM, the existence of a linear hydrostatic term is compatible with the symmetries of the system~\cite{PBS57_main}. A symmetry analysis indeed suggests that, to second order in the strains $\varepsilon_{ij}$ in the cubic axis set:
\begin{align}
A/A_0&=1+\frac{K}{3}(\varepsilon_{xx}+\varepsilon_{yy}+\varepsilon_{zz}) \nonumber \\ 
&+\frac{L}{2}\left[(\varepsilon_{yy}-\varepsilon_{zz})^2+(\varepsilon_{xx}-\varepsilon_{zz})^2+(\varepsilon_{xx}-\varepsilon_{yy})^2\right] \nonumber \\ 
&+N(\varepsilon_{yz}^2+\varepsilon_{xz}^2+\varepsilon_{xy}^2)\,.
\label{eqTB}
\end{align}
A fit to the TB data yields $K=29.3$, $L=-9064$ and $N=-225$. $L$ mostly results from the coupling of the $1s(A_1)$ with the $1s(E)$ state by the uniaxial strain. The TB $L$ is close to the VRM $L=-2\Xi_u^2/(9\Delta^2)=-9720$ \cite{WilsonFeher61}, where $\Xi_u=8.6$ eV is the uniaxial deformation potential of the conduction band of silicon and $\Delta=41$ meV is the splitting between the $1s(A_1)$ and the $1s(E)$ state of the Bi impurity. The quadratic shear term $N$ is usually negligible with respect to $L$. $K=\partial(A/A_0)/\partial\varepsilon_{\rm hs}$ results from the coupling of the $1s(A_1)$ with the $2s(A_1)$ state (and higher $A_1$ states, since hydrostatic strain preserves the symmetry of the system) due to the change of the shape and depth of the central cell correction under strain. $A/A_0$ is dominated by this hydrostatic term at small strain, as evidenced in Fig. \ref{fig:dfda_fits}(B). The TB $K=29.3$ is larger than the experimental $K=19.1$. At variance with $L$ (which mostly depends on a deformation potential of the silicon matrix), $K$ indeed depends on details of the potential near the impurity, which must be specifically accounted for in the TB model in order to reach quantitative accuracy~\cite{suppmat}. In order to better capture the central cell correction around the bismuth impurity, we also performed first principles calculations using density functional theory (DFT) to describe the atomic relaxations not accounted for by our TB calculations~\cite{suppmat}. The DFT calculations further corroborate the linear dependence of the hyperfine coupling on hydrostatic strain (for $\epsilon_{\rm hs} \leq10^{-3}$), and predict a coefficient $K=17.5$,  in good agreement with our experiments. Full details concerning the models and calculations can be found in Ref~\cite{suppmat}. 

\begin{figure}[t]	
	\includegraphics[width=90mm]{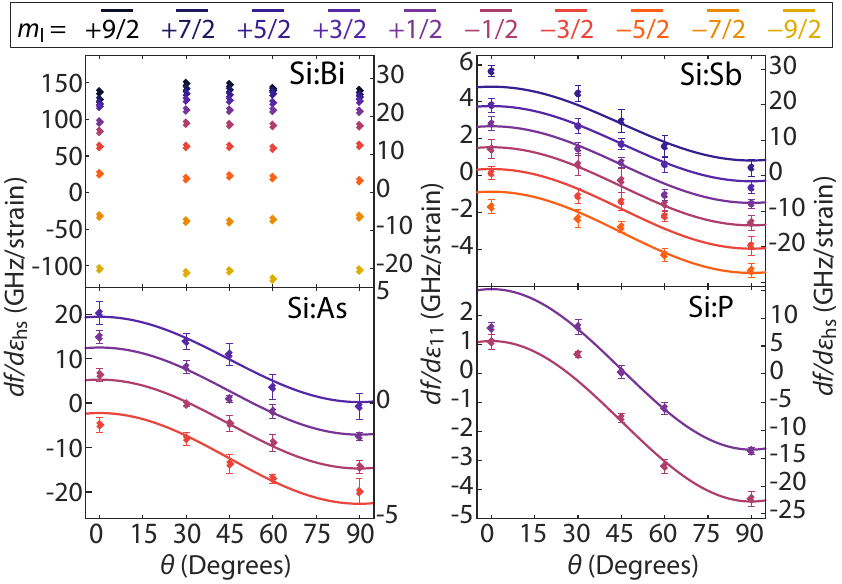}
	\caption{Extracted linear fit gradients d$f/$d$\epsilon_{11}$ for each transition for all four donors under consideration as a function of the angle of $B_0$ w.r.t. the crystal $\theta$. For Si:Sb, Si:As, and Si:P, these fits are overlaid with a model taking into account the linear shift of hyperfine interaction strength $A$ as well as an anisotropic g-factor as a function of $\epsilon_{11}$. \label{fig:anisotropy}}
\end{figure}

To test our model further and explore the expected anisotropy of a g-factor coupling to strain, we extend our study over a range of magnetic field orientations (as defined in Fig.~\ref{fig:figone}(B)) and for the other group V donors:  $^{31}$P, $^{75}$As, and $^{121}$Sb.
In all cases we find the observed ESR transition frequency shifts $f$ are linear as a function of hydrostatic strain $\epsilon_{\rm hs}$, with the resulting coupling strengths ($df/d\epsilon_{\rm hs}$) summarised in Fig.~\ref{fig:anisotropy} and Table~SI, along with values predicted from tight-binding calculations and full data sets~\cite{suppmat}. 
While we find no significant anisotropy in Si:Bi, the data from Si:Sb, Si:As, and Si:P display strain effects which clearly depend on the magnetic field orientation, attributed to a strain-induced anisotropic electronic g-factor. Following Wilson \& Feher \cite{WilsonFeher61}, our model for this anisotropy includes a term accounting for the effect of valley repopulation, and another accounting for the effect of spin-orbit coupling in the sheared lattice. Fits of this model to our experimental data reproduce the predicted strength of both of these effects to within a factor of two~\cite{suppmat}.

Through experiments and calculations, we have demonstrated that hydrostatic strain in silicon leads to a strong, linear tuning of the hyperfine interaction in group V donors, through coupling between the $1s(A_1)$ and $2s(A_1)$ states. The ability to shift the ESR transition frequencies by over a linewidth with hydrostatic strain in the order of $10^{-6}$  opens up new possibilities for conditional ``A-gate'' control of donors as well as coupling to mechanical resonators. In addition, these insights will be crucial in supporting the design of quantum memories and processors based on donors in silicon, enabling the ability to accurately predict ESR transition energies as a function of donor position within the device structure.

We acknowledge helpful discussions with Ania Jayich. This research was supported by the Engineering and Physical Sciences Research Council (EPSRC) through UNDEDD (EP/K025945/1) and a Doctoral Training Grant; as well as by the European Union's Horizon 2020 research and innovation programme under Grant Agreement No 688539 (http://mos-quito.eu) and the European Community's Seventh Framework Programme Nos. 279781 (ASCENT) and 615767 (CIRQUSS); and also by the Agence Nationale de la Recherche through project QIPSE.Ó

\bibliographystyle{apsrev4-1}


\clearpage
\onecolumngrid
\begin{center}
	\textbf{\large Supplementary Materials: Linear hyperfine tuning of donor spins in silicon using hydrostatic strain}
\end{center}
\section{Experimental methods}

\renewcommand{\theequation}{S\arabic{equation}}  
\renewcommand{\thefigure}{S\arabic{figure}} 
\renewcommand{\thetable}{S\Roman{table}}

\setcounter{equation}{0}
\setcounter{figure}{0}
\setcounter{table}{0}
\setcounter{page}{1}
\makeatletter

\renewcommand{\bibnumfmt}[1]{[S#1]}
\renewcommand{\citenumfont}[1]{S#1}

The sample `Bi' is a $2\times2\times10$ mm single crystal of isotopically purified $^{28}$Si doped with \num{4.4e14} Bi donors/cm$^{3}$, and `Buffet' is a $2\times2\times7$ mm $^{28}$Si single crystal doped with \num{1.5e14} $^{31}$P donors/cm$^{3}$, \num{5e14} $^{75}$As donors/cm$^{3}$, and \num{1.1e14} $^{121}$Sb donors/cm$^{3}$. 

\begin{figure}[h]
	\includegraphics{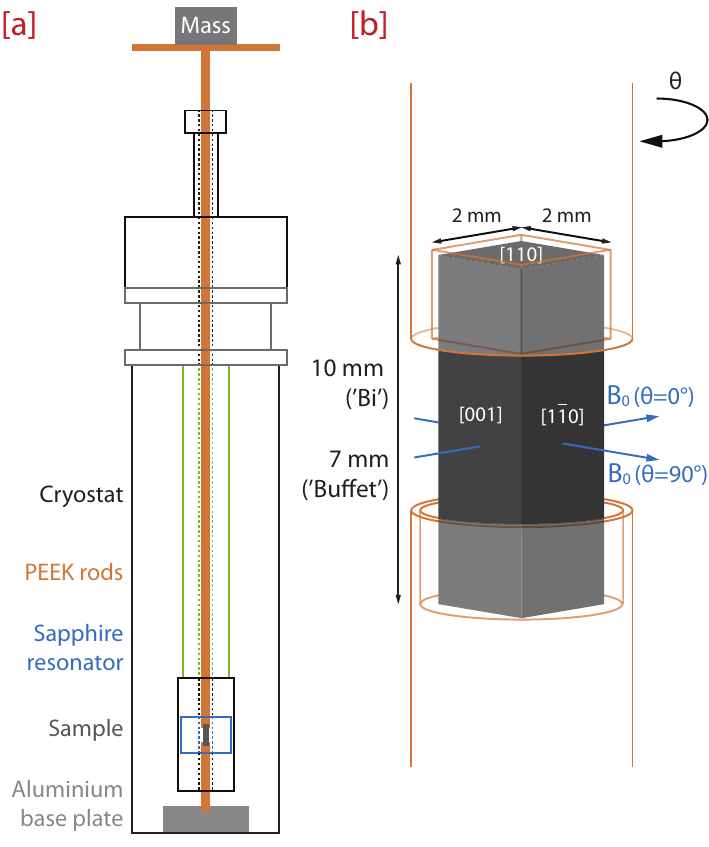}
	\caption{\textbf{a)} Schematic of experimental setup showing Bruker ESR resonator mounted inside liquid Helium flow cryostat. The sample is held in place in the centre of the sapphire resonator by two engineered plastic rods. \textbf{b)} Magnified view of sample inside strain mount.\label{fig:sample}}
\end{figure}

A cylindrical aluminium plate rests on the floor of the cryostat (see Fig. \ref{fig:sample}). A rod made from the plastic PEEK screws into this plate and extends into the centre of the sapphire resonator, providing a bottom support for the sample. A second PEEK rod, which is supported radially by the resonator structure but is free to move along its axial direction, holds the sample in place from above. 

Each echo is averaged 300 times and the spins are reset after each cycle with a 5 ms flash of 50 mW above-bandgap 1047 nm laser light through an optical window in the cryostat. 

The time-domain Hahn echo signals can be expressed as a periodic oscillation at the frequency of the detuning between the fixed microwave drive frequency and the strain-shifted transition frequency, multiplied by an envelope function given by the inverse Fourier transform of the frequency-domain transition spectral lineshape. Then, by the convolution theorem, the Fourier transform of such a signal results in a frequency-domain spectral peak centred at the detuning frequency. The strain-induced detuning is then extracted by fitting a Voigt profile to this spectral peak. See Ref.~\cite{Schweiger05} Chapter 5 for more details. 

\clearpage

\section{Tight-binding modeling}

\subsection{Model}
\label{subsmodel}

We consider a single bismuth impurity at the center of a large box of silicon with side $L=48a\simeq 26$ nm ($a=5.431$\AA\ being the lattice parameter of silicon).

The electronic structure of silicon is described by the $sp^3d^5s^*$ tight-binding (TB) model of Ref. \onlinecite{Niquet09}. This model reproduces the effects of arbitrary strains on the band edges and effectives masses of silicon. Note that this model includes two $s$ orbitals per atom (``$s$'' and ``$s^*$'').

The bismuth impurity is described by a Coulomb tail and an ``on-site'' chemical and Coulomb correction~\cite{Roche12,Usman15}. The expression of the Coulomb tail is based on the dielectric function proposed by Nara~\cite{Nara66}. The potential on atom $i$ reads:
\begin{equation}
V_i=V(R_i)\text{\ with\ }V(r)=-\frac{e^2}{\kappa r}\left(1+A\kappa e^{-\alpha r}+(1-A)\kappa e^{-\beta r}-e^{-\gamma r}\right)\,,
\label{eqVi}
\end{equation}
where $\vec{R}_i$ is the position of the atom (the bismuth impurity being at $\vec{R}_1=\vec{0}$), $\kappa=11.7$ is the dielectric constant of silicon, $A=1.175$, $\alpha=0.757$ Bohrs$^{-1}$, $\beta=0.312$ Bohrs$^{-1}$, and $\gamma=2.044$ Bohrs$^{-1}$~\cite{Nara66,Bernholc77}. This expression deviates from a simple $-e^2/(\kappa R_i)$ tail mostly on the first and second nearest neighbors of the bismuth atom.

\begin{figure}[t]
	\centering
	\includegraphics[width=0.66\columnwidth]{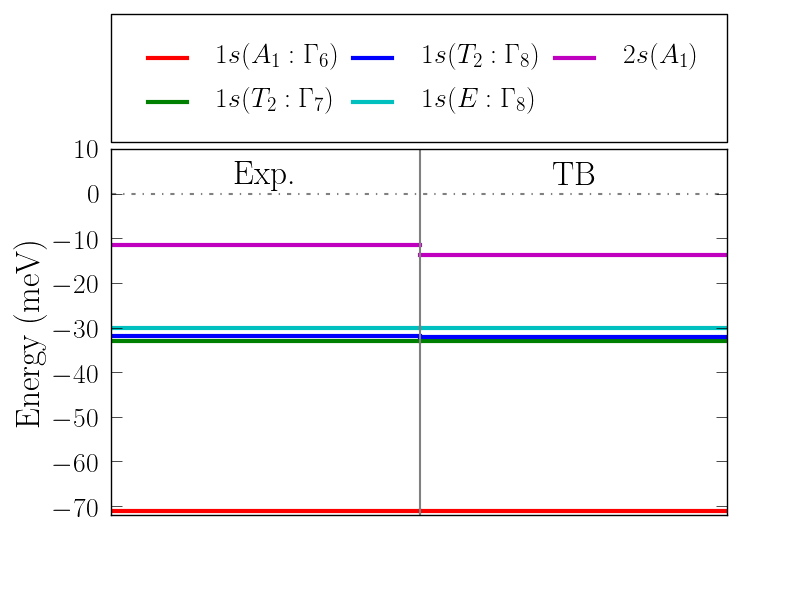}
	\caption{\label{figBi} Experimental (Exp.) and TB bound states of a bismuth impurity in silicon. The horizontal dash-dotted line is the bulk conduction band edge.}
\end{figure}

The ``on-site'' correction is a shift of the energies of the bismuth orbitals that accounts for the different chemical nature of the impurity and for the short-range part of the Coulomb tail. This shift $\Delta E=-U$ reads for each orbital:
\begin{align}
&U_s=5.862\ \rm{eV} \nonumber \\
&U_p=3.690\ \rm{eV} \nonumber \\
&U_d=0.000\ \rm{eV} \nonumber \\
&U_{s^*}=5.862\ \rm{eV}\,.
\label{eqU}
\end{align}
These values were adjusted on the experimental binding energies of the $1s(A_1)$, $1s(E)$ and $1s(T_2)$ states of bismuth in silicon (see Fig. \ref{figBi})~\cite{Krag70,Ramdas81,Zhukavin11}. We have set $U_s=U_{s^*}$ on purpose since it is practically difficult to adjust $U_s$ and $U_{s^*}$ separately. We have designed an other model with $U_{s^*}=0$ that gives very similar results. 

We include spin-orbit coupling (SOC) in the calculations. SOC is described by an intra-atomic Hamiltonian acting on the $p$ orbitals of each atom, $H_{\rm SO}=\lambda\vec{L}_i\cdot\vec{S}$, where $\vec{S}$ is the spin, $\vec{L}_i$ the angular momentum on atom $i$, $\lambda=0.0185$ eV for silicon, and $\lambda=0.350$ eV for bismuth (adjusted on the experimental spin splittings of bismuth in silicon~\cite{Zhukavin11}).

The hyperfine coupling constant $A$ is proportional to the probability of presence $|\Psi(\vec{0})|^2$ of the electron on the bismuth nucleus [Eq. (2) of main text]. In the TB framework,
\begin{equation}
\left|\Psi(\vec{0})\right|^2=\left|c_s s(\vec{0})+c_{s^*} s^*(\vec{0})\right|^2=\left|c_s\right|^2\left|s(\vec{0})\right|^2\left|1+\frac{c_{s^*}}{c_s}\frac{s^*(\vec{0})}{s(\vec{0})}\right|^2\,,
\end{equation}
where $c_s$ and $c_{s^*}$ are the coefficients of the $s$ and $s^*$ orbitals of the bismuth atom in the TB wavefunctions (we discard the spin index here for the sake of simplicity). This expression is, in principle, ambiguous because the radial parts of the $s$ and $s^*$ orbitals of the TB model are not explicitly known. We have tentatively set $R_{s^*s}=s^*(\vec{0})/s(\vec{0})=0.058$~\cite{Usman15}. Yet the choice for $R_{s^*s}$ is practically little relevant, as $c_{s^*}/c_s$ is almost independent on the strains $\hat{\varepsilon}=\{\varepsilon_{xx}, \varepsilon_{yy}, \varepsilon_{zz}, \varepsilon_{yz}, \varepsilon_{xz}, \varepsilon_{xy}\}$. Therefore, the quantity
\begin{equation}
{\cal A}(\hat{\varepsilon})=\frac{A(\hat{\varepsilon})}{A_0}=\frac{\left|\Psi(\vec{0},\hat{\varepsilon})\right|^2}{\left|\Psi(\vec{0}, \hat{\varepsilon}=0)\right|^2}\,,
\end{equation}
which describes the relative change of the hyperfine coupling constant under strains, is well defined within TB, irrespective of the assumptions made for the radial parts of the $s$ and $s^*$ orbitals.

\newpage

\subsection{Strains}

We consider uniaxial stress along $[001]$ and $[110]$.

For uniaxial stress $\sigma_{zz}=\sigma_\parallel$ along $[001]$, the infinitesimal strains in the cubic axis set can be found from Hooke's law $\sigma_{xx}=\sigma_{yy}=0$:
\begin{subequations}
	\label{eq001stress}
	\begin{align}
	\varepsilon_{zz}&=\varepsilon_\parallel=\frac{c_{11}+c_{12}}{(c_{11}-c_{12})(c_{11}+2c_{12})}\sigma_\parallel \\ 
	\varepsilon_{xx}=\varepsilon_{yy}&=\varepsilon_\perp=-\frac{c_{12}}{c_{11}+c_{12}}\varepsilon_\parallel\,,
	\end{align}
\end{subequations}
where $c_{11}=166$ GPa, $c_{12}=64$ GPa and $c_{44}=79.6$ GPa are the elastic constants of bulk silicon.

For uniaxial stress along $[110]$, the infinitesimal strains in the $\{1\equiv[110],\,2\equiv[1\bar{1}0],\,3\equiv[001]\}$ axis set read:
\begin{subequations}
	\label{eq110stress}
	\begin{align}
	\varepsilon_{11}=\varepsilon_\parallel&=\frac{(c_{11}-c_{12})(c_{11}+2c_{12})+2c_{11}c_{44}}{4(c_{11}-c_{12})(c_{11}+2c_{12})c_{44}}\sigma_\parallel \\
	\varepsilon_{22}&=-\frac{(c_{11}-c_{12})(c_{11}+2c_{12})-2c_{11}c_{44}}{(c_{11}-c_{12})(c_{11}+2c_{12})+2c_{11}c_{44}}\varepsilon_\parallel \\		
	\varepsilon_{33}&=-\frac{4c_{12}c_{44}}{(c_{11}-c_{12})(c_{11}+2c_{12})+2c_{11}c_{44}}\varepsilon_\parallel\,.
	\end{align}
\end{subequations}
In the original cubic axis set, the strains are therefore:
\begin{subequations}
	\begin{align}
	\varepsilon_{xx}=\varepsilon_{yy}&=\frac{2c_{11}c_{44}}{(c_{11}-c_{12})(c_{11}+2c_{12})+2c_{11}c_{44}}\varepsilon_\parallel \\
	\varepsilon_{zz}&=-\frac{4c_{12}c_{44}}{(c_{11}-c_{12})(c_{11}+2c_{12})+2c_{11}c_{44}}\varepsilon_\parallel \\
	\varepsilon_{xy}&=\frac{(c_{11}-c_{12})(c_{11}+2c_{12})}{(c_{11}-c_{12})(c_{11}+2c_{12})+2c_{11}c_{44}}\varepsilon_\parallel\,.
	\end{align}
\end{subequations}
Note that there is an additional shear component with respect to uniaxial $[001]$ stress. 

\subsection{Results}

\begin{figure}[!t]
	\centering
	\includegraphics[width=0.48\columnwidth]{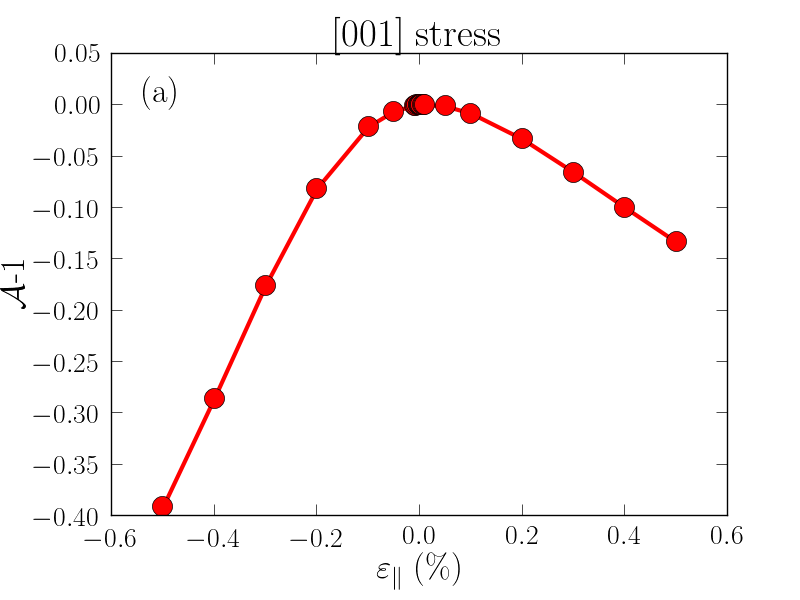}
	\includegraphics[width=0.48\columnwidth]{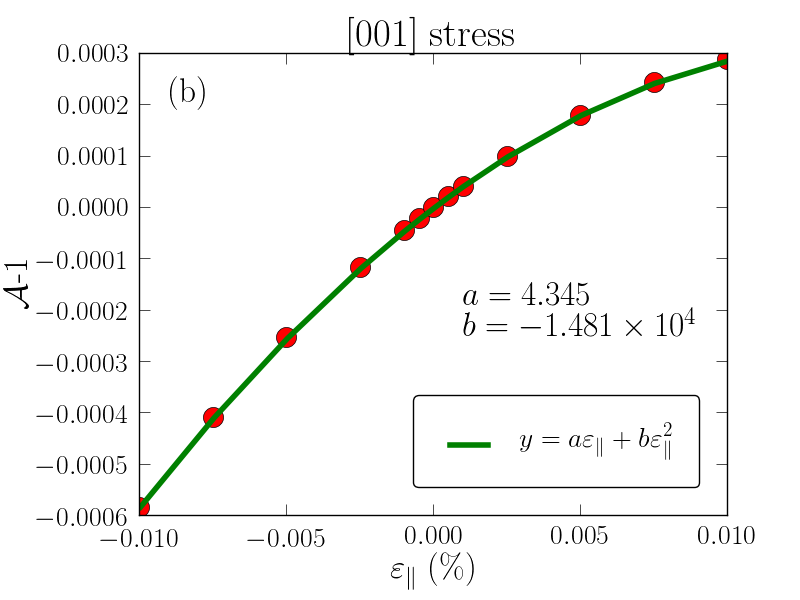} \\
	\includegraphics[width=0.48\columnwidth]{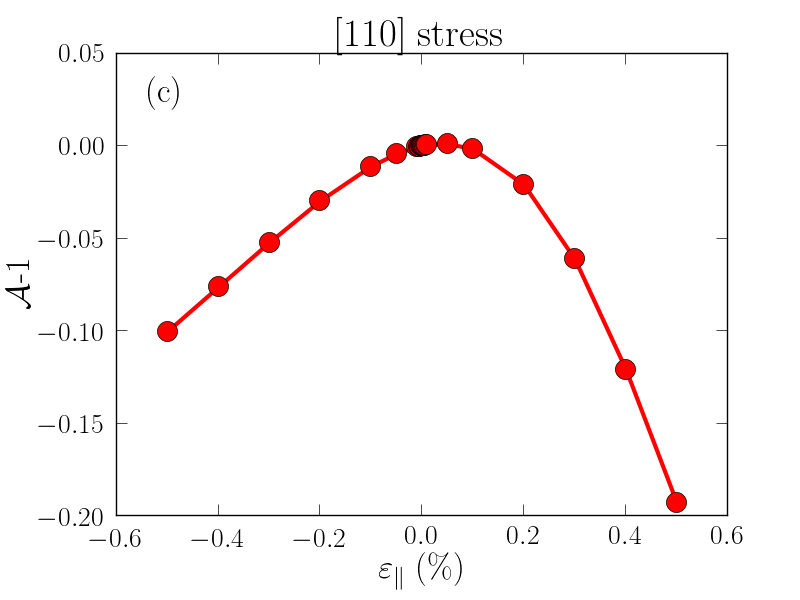}
	\includegraphics[width=0.48\columnwidth]{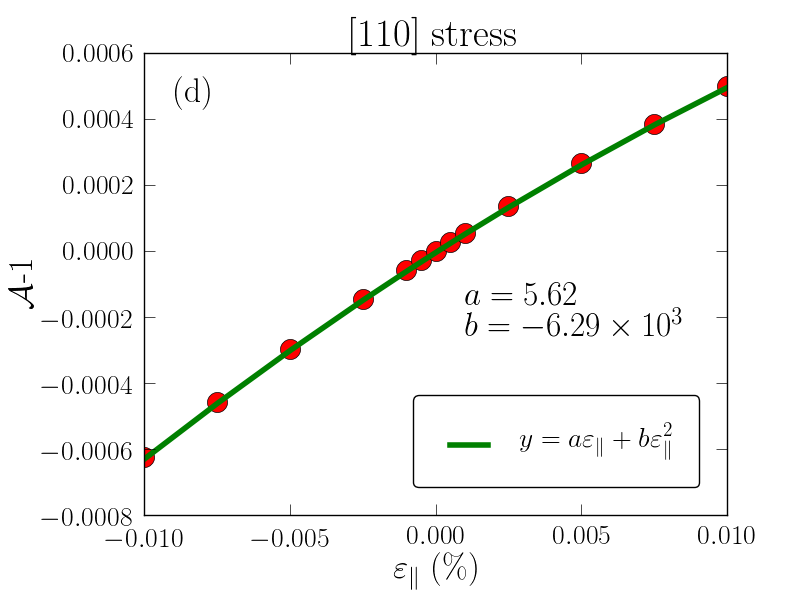} \\
	\caption{\label{figAstress} (a, b) ${\cal A}(\varepsilon_\parallel)$ for uniaxial $[001]$ stress on (a) large and (b) small scales. (c, d) ${\cal A}(\varepsilon_\parallel)$ for uniaxial $[110]$ stress on (c) large and (d) small scales.}
\end{figure}

${\cal A}(\varepsilon_\parallel)$ is plotted in Fig. \ref{figAstress} for uniaxial $[001]$ and $[110]$ stress.

In the valley repopulation model (VRM)~\cite{Feher61}, ${\cal A}(\varepsilon_\parallel)$ is expected to be quadratic with small $\varepsilon_\parallel$ (the changes in ${\cal A}$ being exclusively driven by the loss of symmetries). In the TB approximation, ${\cal A}(\varepsilon_\parallel)$ indeed describes a parabola for weak stress, but centered on some $\varepsilon_\parallel>0$. Therefore, ${\cal A}(\varepsilon_\parallel)$ appears to behave almost linearly with small compressive $\varepsilon_\parallel$.

To understand this trend, it is very instructive to split the strain into a hydrostatic component, an uniaxial component, and a shear component. The hydrostatic component $\hat{\varepsilon}_{\rm hs}$ is defined as $\varepsilon_{xx}=\varepsilon_{yy}=\varepsilon_{zz}=\varepsilon_{\rm hs}$. It accounts for the changes in the total volume $\Omega$ ($\Delta\Omega/\Omega=3\varepsilon_{\rm hs}$). The uniaxial and shear components account for the changes in symmetries (at constant volume). The uniaxial component $\hat{\varepsilon}_{\rm uni}$ is defined as $\varepsilon_{zz}=\varepsilon_{\rm uni}$, $\varepsilon_{xx}=\varepsilon_{yy}=-\varepsilon_{\rm uni}/2$, and the shear component $\hat{\varepsilon}_{\rm shear}$ as $\varepsilon_{xy}=\varepsilon_{\rm shear}$.

For uniaxial stress along $[001]$ [Eqs. (\ref{eq001stress})], 
\begin{subequations}
	\begin{align}
	\varepsilon_{\rm hs}^{001}&=\frac{k}{3}\varepsilon_\parallel \\
	\varepsilon_{\rm uni}^{001}&=\left(1-\frac{k}{3}\right)\varepsilon_\parallel \,,
	\end{align}
\end{subequations}
where:
\begin{equation}
k=\frac{c_{11}-c_{12}}{c_{11}+c_{12}}\,. 
\end{equation}

For uniaxial stress along $[110]$ [Eqs. (\ref{eq110stress})],
\begin{subequations}
	\begin{align}
	\varepsilon_{\rm hs}^{110}&=\frac{2k_{1}-k_{2}}{3}\varepsilon_\parallel  \\
	\varepsilon_{\rm uni}^{110}&=-\frac{2(k_{1}+k_{2})}{3}\varepsilon_\parallel \\
	\varepsilon_{\rm shear}^{110}&=(1-k_1)\varepsilon_\parallel
	\end{align}
\end{subequations}
where:
\begin{subequations}
	\label{eq:k1k2}
	\begin{align}
	k_{1}&=\frac{2c_{11}c_{44}}{(c_{11}-c_{12})(c_{11}+2c_{12})+2c_{11}c_{44}} \\
	k_{2}&=\frac{4c_{12}c_{44}}{(c_{11}-c_{12})(c_{11}+2c_{12})+2c_{11}c_{44}} \,.
	\end{align}
\end{subequations}

\begin{figure}[!t]
	\centering
	\includegraphics[width=0.66\columnwidth]{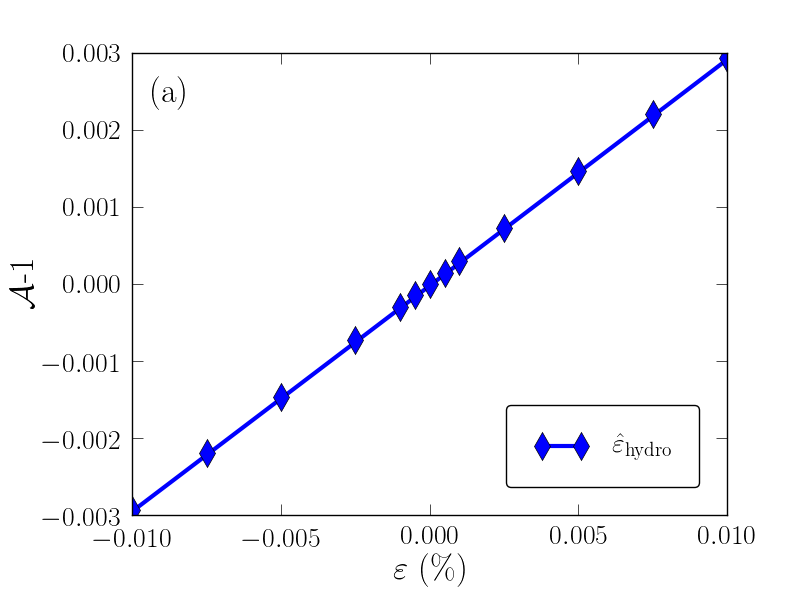}
	\includegraphics[width=0.66\columnwidth]{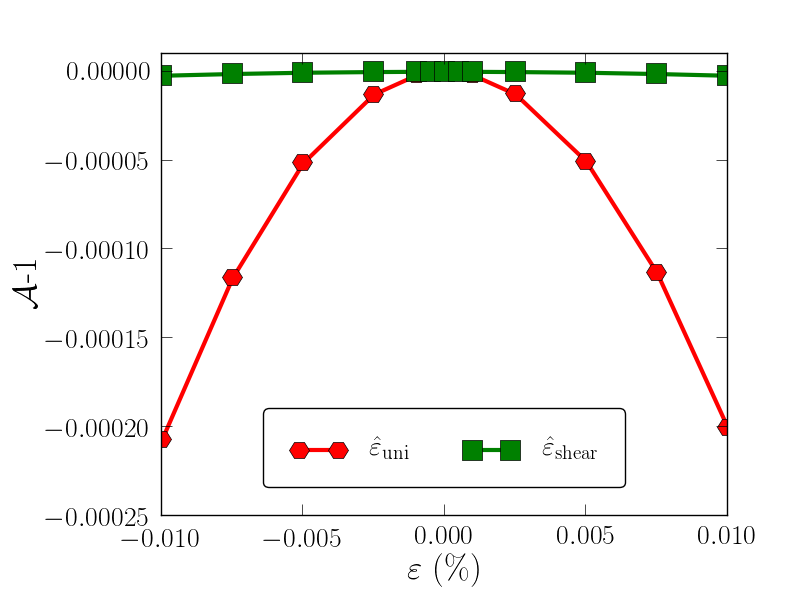}
	\caption{\label{figAcomponents} ${\cal A}(\hat{\varepsilon})$ for (a) hydrostatic strain ($\varepsilon=\varepsilon_{\rm hs}$), (b) uniaxial ($\varepsilon=\varepsilon_{\rm uni}$) and shear strain ($\varepsilon=\varepsilon_{\rm shear}$).}
\end{figure}

${\cal A}(\hat{\varepsilon})$ is plotted as a function of $\hat{\varepsilon}_{\rm hs}$, $\hat{\varepsilon}_{\rm uni}$,  and $\hat{\varepsilon}_{\rm shear}$ in Fig. \ref{figAcomponents}. ${\cal A}(\hat{\varepsilon})$ shows a quadratic behavior as a function of $\varepsilon_{\rm uni}$ and $\varepsilon_{\rm shear}$. It does, however, behave linearly as a function of $\varepsilon_{\rm hs}$. The trends at small $\varepsilon_\parallel<0$ evidenced in Fig. \ref{figAstress} can, therefore, be ascribed to the effects of hydrostatic strains on the hyperfine coupling constant. As a matter of fact, ${\cal A}(\varepsilon_\parallel)={\cal A}(\hat{\varepsilon}_{\rm hs})+{\cal A}(\hat{\varepsilon}_{\rm uni}^{001})$ for uniaxial $[001]$ stress, and ${\cal A}(\varepsilon_\parallel)={\cal A}(\hat{\varepsilon}_{\rm hs})+{\cal A}(\hat{\varepsilon}_{\rm uni}^{110})+{\cal A}(\hat{\varepsilon}_{\rm shear}^{110})$ for uniaxial $[110]$ stress, showing the relevance of this decomposition.

\subsection{Discussion}

Although not predicted by the valley repopulation model, the existence of a $\propto\varepsilon_{\rm hs}$ term in ${\cal A}$ is allowed by symmetries~\cite{PBS57}. Indeed, a symmetry analysis suggests that, to second order in $\varepsilon_{ij}$:
\begin{align}
{\cal A}-1&=\frac{K}{3}(\varepsilon_{xx}+\varepsilon_{yy}+\varepsilon_{zz}) \nonumber \\ 
&+L(\varepsilon_{xx}^2+\varepsilon_{yy}^2+\varepsilon_{zz}^2)+M(\varepsilon_{yy}\varepsilon_{zz}+\varepsilon_{xx}\varepsilon_{zz}+\varepsilon_{xx}\varepsilon_{yy})+N(\varepsilon_{yz}^2+\varepsilon_{xz}^2+\varepsilon_{xy}^2)\,,
\label{eqA1}
\end{align}
where $K=\partial{\cal A}/\partial\varepsilon_{\rm hs}$, $L$, $M$ and $N$ are constants. A fit to the tight-binding data yields:
\begin{align}
K&=29.3 \nonumber \\
L\simeq-M&=-9064 \nonumber \\
N&=-225 
\end{align}
Eq. (\ref{eqA1}) then simplifies into:
\begin{align}
{\cal A}-1&=\frac{K}{3}(\varepsilon_{xx}+\varepsilon_{yy}+\varepsilon_{zz}) \nonumber \\ 
&+\frac{L}{2}\left[(\varepsilon_{yy}-\varepsilon_{zz})^2+(\varepsilon_{xx}-\varepsilon_{zz})^2+(\varepsilon_{xx}-\varepsilon_{yy})^2\right]+N(\varepsilon_{yz}^2+\varepsilon_{xz}^2+\varepsilon_{xy}^2)\,.
\label{eqA2}
\end{align}
We find $L\simeq-M$ because there is no sizable non-linearity in the dependence of ${\cal A}$ on hydrostatic strain in the investigated range $|\varepsilon_{\rm hs}|<10^{-4}$ (no $\propto\varepsilon_{\rm hs}^2$ term above when $\varepsilon_{xx}=\varepsilon_{yy}=\varepsilon_{zz}$). 
Also note that the effects of the quadratic shear terms are usually negligible with respect to the effects of the quadratic uniaxial terms ($N\ll L$).

The quadratic $L$ term is mostly due to the coupling of the $1s(A_1)$ with the $1s(E)$ states of the impurity under uniaxial strain. The VRM of Ref. \onlinecite{Feher61} actually suggests $L=-2\Xi_u^2/(9\Delta^2)$, where $\Xi_u=8.6$ eV is the uniaxial deformation potential of the conduction band of silicon and $\Delta$ is the splitting between the $1s(A_1)$ and the $1s(E)$ state. For Bi ($\Delta=41$ meV), the VRM predicts $L=-9720$, in close agreement with the TB data. The linear hydrostatic term is due, on the other hand, to the coupling of the $1s(A_1)$ with $2s(A_1)$ state (and possibly higher $s$ states with the same symmetry). This coupling results from the variations of the on-site correction on the bismuth impurity, and from the variations of the bismuth-silicon interactions under hydrostatic strain -- in other words, from the variations of the depth and shape of the ``central cell correction''~\cite{Lipari80} not accounted for by the VRM.

Indeed, the total potential on the bismuth atom catches contributions from the tails of the atomic potentials of the neighboring silicon atoms, and these contributions depend on the silicon-bismuth bond lengths. As a matter of fact, the present TB model includes a strain-dependent correction for the energy $E_{i\mu}$ of orbital $\mu\equiv s,p,d,s^*$ of atom $i$~\cite{Niquet09}:
\begin{equation}
E_{i\mu}=E_{i\mu}^0+\frac{3}{4}\alpha_{i\mu}\sum_{j\in{\rm NN}(i)}\frac{d_{ij}-d_{ij}^0}{d_{ij}^0}=E_{i\mu}^0+3\alpha_{i\mu}\varepsilon_{\rm hs}\,,
\end{equation}
where the sum runs over the nearest neighbors $j$ of atom $i$, $d_{ij}$ is the distance between atoms $i$ and $j$, and $d_{ij}^0$ is the relaxed bond length. $E_{i\mu}^0$ is the energy of the orbital in the reference, unstrained system and $\alpha_{i\mu}$ characterizes the deepening of the potential under strain. The interactions between bismuth and the nearest neighbor silicon atoms scale, on the other hand, as $(d_{ij}/d_{ij}^0)^n$, with $n$ close to 2. 

In the present TB model, the parameters $\alpha_{i\mu}$ and the exponents $n$ of bismuth are the same as for silicon. Although this choice is a safe first guess, it can only provide a semi-quantitative description of the dependence of ${\cal A}$ on hydrostatic strain. This is why the TB $K=29.3$ is significantly larger than the experimental $K=19.1$. We may, in the spirit of Eqs. (\ref{eqU}), lump all corrections to the TB model into the $\alpha_{i\mu}$. Therefore, we tentatively set:
\begin{equation}
\alpha_{i\mu}({\rm Bi})=\alpha_{i\mu}({\rm Si})+\Delta\alpha\,,
\label{eqdalpha}
\end{equation}
and adjust $\Delta\alpha$ on the experimental $K$. This yields $\Delta\alpha=3.48$ eV for bismuth. 

\begin{table}
	\begin{tabular}{l|r|ddd|d|drr}
		& Coulomb tail & \multicolumn{1}{r}{$U_s=U_{s^*}$ (eV)} & \multicolumn{1}{r}{$U_p$ (eV)} & \multicolumn{1}{r|}{$U_d$ (eV)} & \multicolumn{1}{r|}{$\Delta\alpha$ (eV)} & \multicolumn{1}{r}{$K$} & \multicolumn{1}{r}{$L$} & \multicolumn{1}{r}{$N$} \\
		\hline
		P & $-e^2/(\kappa R_i)$ & 4.535 & 2.405 & 2.055 & -19.50 & 79.5 & -103640 & 1277 \\
		As & $-e^2/(\kappa R_i)$ & 5.060 & 2.330 & 0.325 & -2.68 & 37.2 & -33836 & 833 \\
		Sb & Nara [Eq. (\ref{eqVi})] & 4.629 & 4.448 & 0.000 & -2.54 & 32.6 & -104340 & 1420 \\
		Bi & Nara [Eq. (\ref{eqVi})] & 5.862 & 3.690 & 0.000 & 3.48 & 19.1 & -9064 & -225
	\end{tabular}
	\caption{Nature of the Coulomb tail, on-site corrections $U$ [Eq. (\ref{eqU})] and $\Delta\alpha$ [Eq. (\ref{eqdalpha})], and value of $K$, $L$ and $N$ for P, As, Sb and Bi donors in silicon.}
	\label{tableimp}
\end{table}

We have repeated the same procedure for P, As and Sb. We give in Table \ref{tableimp} the on-site parameters $U_s=U_{s^*}$, $U_p$, $U_d$ and $\Delta\alpha$ of each impurity, as well as the values of $K$, $L$ and $N$. The model for P and As~\cite{Diarra07} is based on a simple Coulomb tail $V(R_i)=-e^2/(\kappa R_i)$ instead of Eq. (\ref{eqVi}). 

The dependence of the binding energy $E_b$ of As impurities on the hydrostatic pressure $P$ has been measured by Holland and Paul ($dE_b/dP\simeq-0.05$ meV/kbar)~\cite{HollandPaul62} and by Samara and Barnes ($dE_b/dP\simeq-0.1$ meV/kbar)~\cite{SamaraBarnes87}. The electron hence gets more loosely bound to the impurity under pressure (or equivalently under compressive hydrostatic strain). This is consistent with the decrease of the hyperfine coupling constant $A$ reported here ($K<0$).
Samara and Barnes explain the decrease of $E_b$ under pressure by the variations of the effective masses and dielectric constant $\kappa$. We point out, though, that there is also a significant contribution from the variations of the central cell correction. The variations of effective masses and dielectric constant actually make little contribution to $K$. In the simplest effective mass approximation, the wave function of the electron bound to the donor is indeed $\Psi(\vec{r})=e^{-r/a_B}/(\sqrt{\pi}a_B^{3/2})$, where $a_B=\hbar^2\kappa/(m^*e^2)$ is the Bohr radius. Hence, $|\Psi(\vec{0})|^2=1/(\pi a_B^3)$, so that:
\begin{equation}
K=3\left(\frac{1}{m^*}\frac{\partial m^*}{\partial\varepsilon_{\rm hs}}-\frac{1}{\kappa}\frac{\partial\kappa}{\partial\varepsilon_{\rm hs}}\right)
\end{equation}
{\it Ab-initio} calculations within density functional theory (see next section) give $(1/\kappa)(\partial\kappa/\partial\varepsilon_{\rm hs})=0.78$, $(1/m^*)(\partial m^*/\partial\varepsilon_{\rm hs})=-0.19$ for the longitudinal mass, and $(1/m^*)(\partial m^*/\partial\varepsilon_{\rm hs})=1.54$ for the transverse mass. Therefore, the variations of the masses and dielectric constant are expected to have little net effect on the hyperfine coupling constant.

\section{Density functional theory calculations}

In order to strengthen the above interpretation, we have also performed first principles calculations using density functional theory (DFT) with the Perdew-Burke-Ernzerhof (PBE) exchange-correlation functional~\cite{Perdew96} and the projector-augmented wave method~\cite{Blochl94} in the Vienna Ab-initio Simulation Package (VASP)~\cite{Kresse96}, following the methodology described in Ref.~\cite{Blochl00}. The calculations were carried out on one Bi impurity in a 1728-atom supercell, with a 250 eV plane-wave cutoff energy. DFT describes the central cell correction around the bismuth impurity from first principles and captures the atomic relaxations not accounted for by TB calculations --- due to its accuracy in the immediate vicinity of the donor it could be expected to provide a good description of the variation of the hyperfine coupling with strains. Nevertheless, due to the finite size of the supercell, DFT misses the long range Coulomb tail of the potential~\cite{Niquet10}, which (along with over-delocalization arising from the self-interaction error in PBE) contributes to an significant underestimation in the absolute value the hyperfine interaction (1102~MHz). 

The {\it ab-initio} hyperfine coupling shows the expected linear dependence on hydrostatic strain over the entire range explored here (up to $\varepsilon_{xx}=\varepsilon_{yy}=\varepsilon_{zz}=2\times 10^{-3}$). We extract a coefficient $K=17.5$, in good agreement with the experimental data. The {\it ab-initio} quadratic term is $L=-1.17\times 10^4$, as obtained from a fit to the calculated data with $\varepsilon_{zz}\leq 10^{-3}$ and $\varepsilon_{xx}=\varepsilon_{yy} = 0$. Deviations are observed for higher strains, consistent with the higher-order terms present in the VRM. The equilibrium bismuth-silicon bond length (2.651 {\AA}) is significantly larger than the Si-Si bond length (2.367 {\AA} in the bulk). All Bi-Si and Si-Si bonds are simply scaled by the hydrostatic strain, to within better than 0.001 {\AA}. 

These data provide further support for the linear dependence of the hyperfine parameter on the hydrostatic component of strain, and illustrate the complementary strengths of these two computational approaches (DFT and tight-binding) in modelling the behaviour of donors in silicon. 


\section{Modeling g-factor anisotropy}

The ellipsoidal shape of the Si conduction band minima in $k$-space results in differing effective masses for Bloch waves parallel and perpendicular to the valley axis at these points~\cite{YuCardona99}. This leads to a single-valley g-factor which is anisotropic, with $g_\mu^2 = g_\parallel^2\cos^2\phi  + g_\perp^2\sin^2\phi $, where $g_\parallel$ and $g_\perp$ are the g-factors parallel and perpendicular to the valley axis, and $\phi$ is the angle between the magnetic field and the valley axis. In the case of a donor, the g-factor is found by summing over the relative contribution from each valley such that $g_{\rm donor} = \sum_{\mu=1}^6\alpha_\mu^2g_\mu$. For the unstrained $1s(A_1)$ donor ground state, which is an equal superposition of all six equivalent valleys, this summation leads to a cancellation of the anisotropy, leaving $g_0 = \frac{1}{3}g_\parallel + \frac{2}{3}g_\perp$. Under strain, the valleys repopulate, breaking this cancellation symmetry. In our system with $\theta$ defined as in figure \ref{fig:sample}, the resulting g-factor anisotropy can be modeled using the VRM~\cite{WilsonFeher61} by: 
\begin{equation}
\left.\cfrac{\partial g}{\partial\epsilon_{11}}\right\rvert_{\textrm{VRM}}=\beta_{\textrm{VRM}}(3\cos^2\theta-1)
\end{equation}
where:
\begin{equation}
\beta_{\textrm{VRM}}=\cfrac{2\Xi_u}{9\Delta}(g_\parallel-g_\perp)(k_1+k_2)\,.
\label{eq:gvrm}
\end{equation}
$g_\parallel$ and $g_\perp$ are the parallel and perpendicular g-factors, $\Xi_u$ is the uniaxial deformation potential, $\Delta$ is the donor-dependent $1s(A_1)$-$1s(E)$ splitting, and $k_1$ and $k_2$ are defined in Eq.~\ref{eq:k1k2}.

It is also known that the g-factor of a single valley is changed in the presence of shear strain by the coupling with the opposite valley. Following Wilson \& Feher~\cite{WilsonFeher61}, the effective Hamiltonian for this mechanism can be written~\cite{YMNshearnote}:

\begin{equation}
\mathcal{\hat{H}}_{\textrm{shear}}=C\mu_B\varepsilon_{xy}(B_xS_y + B_yS_x)
\label{eqHshear}
\end{equation}
where $C$ is a coefficient involving spin-orbit coupling matrix elements~\cite{WilsonFeher61}. Rewriting the corresponding g-factor contribution in terms of $\varepsilon_{11}$ results in a second anisotropic term:
\begin{equation}
\left.\cfrac{\partial g}{\partial\epsilon_{11}}\right\rvert_{\textrm{shear}} = \beta_{\textrm{shear}}(3\cos^2\theta - 3)
\end{equation}
where:
\begin{equation}
\beta_{\textrm{shear}}=\cfrac{C(1-k_1)}{3}\,.
\label{eq:gshear}
\end{equation}

Then, the derivative of the ESR transition frequency with respect to the uniaxial strain reads:
\begin{equation}
\cfrac{\textrm{d}f}{\textrm{d}\varepsilon_{11}} = \cfrac{\partial A}{\partial \varepsilon_{11}}\cfrac{\partial f}{\partial A} +  \Bigg[\left.\cfrac{\partial g}{\partial\epsilon_{11}}\right\rvert_{\textrm{VRM}} + \left.\cfrac{\partial g}{\partial\epsilon_{11}}\right\rvert_{\textrm{shear}}\Bigg]\cfrac{\partial f}{\partial g}\,.
\end{equation}
$\partial f/\partial g$ can be calculated for each transition in the same manner as $\partial f/\partial A$ by solving the spin Hamiltonian while varying the value of g. We use a linear least squares regression to fit this model to the experimental data. We introduce three fitting parameters characterising the strength of the different effects for each donor: $K=\partial (A/A_0)/\partial \epsilon_{\rm hs}$, $\beta_{\textrm{VRM}}$ and $\beta_{\textrm{shear}}$. The results of these fits are reported in table \ref{tab:summary} and figure 4 in the main text for Si:Sb, Si:As, and Si:P. The absence of a clear anisotropy for Si:Bi could be explained by the relatively small magnitude of the predicted g-factor effects in comparison with the absolute shifts due to the modified hyperfine interaction. 

\begin{center}
	\begin{table}
		\renewcommand\arraystretch{1.2}
		\begin{tabular}{| r | c | c | c | c | c |}
			\hline
			\thead{Donor} &\thead{$K=\frac{\partial (A/A_0)}{\partial \epsilon_{\rm hs}}$ \\(exp.)} & \thead{$\beta_{\textrm{VRM}}\times 10^{-3}$\\ (theory) }& \thead{$\beta_{\textrm{VRM}}\times 10^{-3}$\\ (exp.)} & \thead{$\beta_{\textrm{shear}}\times 10^{-3}$\\ (theory)} & \thead{$\beta_{\textrm{shear}}\times 10^{-3}$ \\(exp.)}\\ \hline
			
			$^{31}$P & $79.2\pm25.7$ & $126.7$ & $206\pm51$& $78$ & $173\pm33$   \\ \hline
			
			$^{75}$As & $37.4\pm 3.3$ & $77.5$ & $165.1\pm24.6$& $78$ & $96\pm15.6$  \\ \hline
			
			$^{121}$Sb & $32.8\pm 2.3$ & $146.0$ & $197.6\pm25.8$& $78$ & $90.3\pm16.4$   \\ \hline
			
			$^{209}$Bi &  $19.1\pm0.3$ & $42.5$ & - & $78$ & - \\ \hline
			
		\end{tabular}
		\caption{Parameters $K=\partial (A/A_0)/\partial \epsilon_{\rm hs}$, $\beta_{\textrm{VRM}}$ and $\beta_{\textrm{shear}}$ extracted from the experimental data, along with theoretical values for $\beta_{\textrm{VRM}}$ and $\beta_{\textrm{shear}}$. The theoretical values are calculated using $C=0.44$~\cite{WilsonFeher61}, $\Delta$ values from Ref.~\onlinecite{Ramdas81}, and $(g_\parallel-g_\perp)$ values from Ref.~\cite{WilsonFeher61}. We assume $(g_\parallel-g_\perp)\approx 1.1\times10^{-3}$ for Si:Bi.}\label{tab:summary}
	\end{table}
\end{center}

We compare the extracted parameters with theoretical predictions for $\beta_{\textrm{VRM}}$ and $\beta_{\textrm{shear}}$ [Eqs. (\ref{eq:gvrm}) and (\ref{eq:gshear})] calculated using $C=0.44$~\cite{WilsonFeher61}, $\Delta$ values from Ref.~\onlinecite{Ramdas81}, and $g_\parallel-g_\perp$ values from Ref.~\cite{WilsonFeher61}. The measured strengths of the g-factor effects agree with theory within approximately a factor of 2. Tight binding simulations with the present model (see also Ref.~\onlinecite{Rahman09}) predict that the very small $g_\parallel-g_\perp$ is approximately an order of magnitude larger than given by Ref.~\cite{WilsonFeher61}. It is interesting to note that the experimental data sit between Ref.~\cite{WilsonFeher61} and the TB predictions.

\clearpage

\section{Full Experimental dataset for all donors}

\begin{figure}[h]	
	\includegraphics{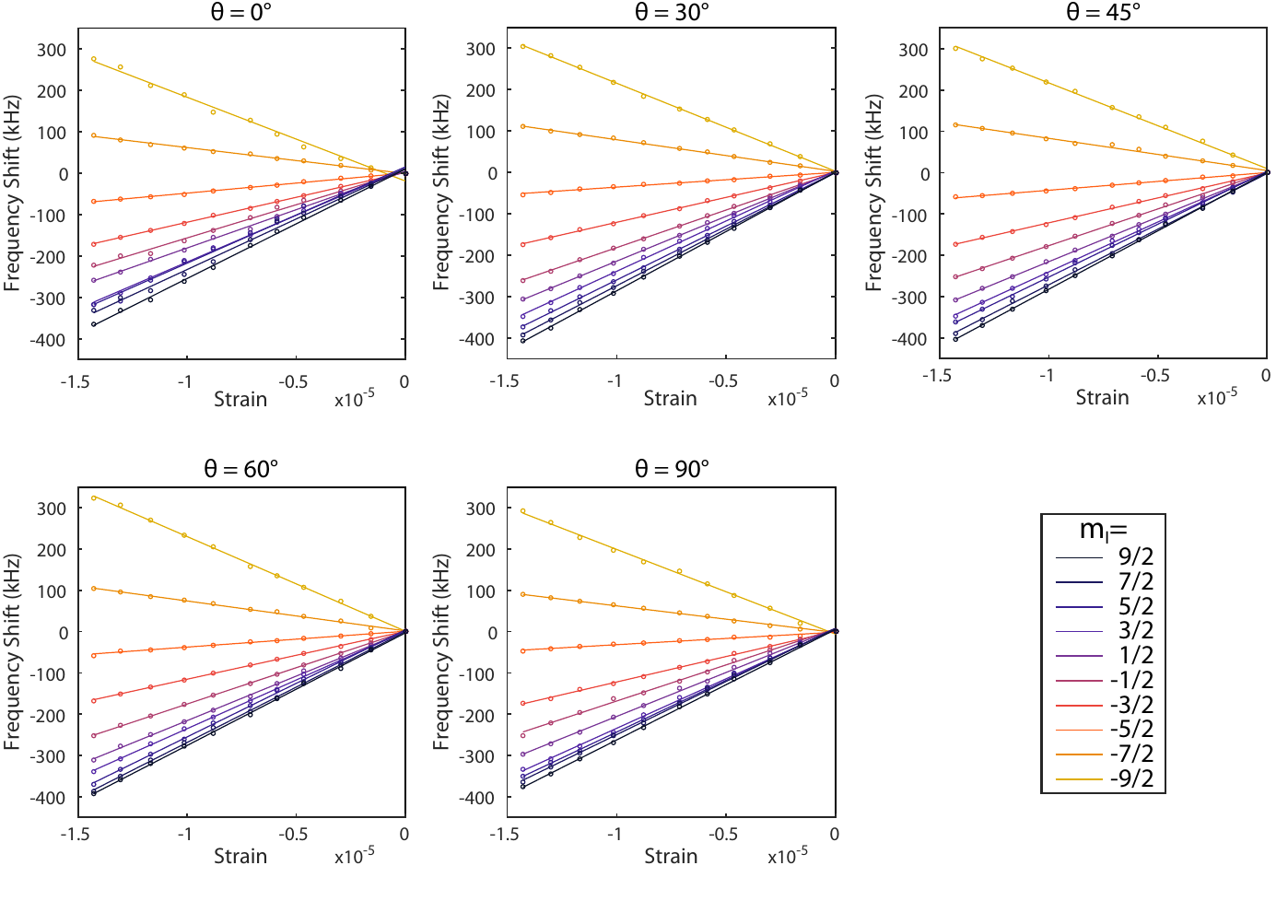}
	\caption{Full dataset for Si:Bi showing frequency shifts for all ten ESR transitions as a function of strain $\varepsilon_{11}$ and $\theta$. Linear fits are shown for each transition. \label{fig:SiBi_full}}
\end{figure}

\begin{figure}[h]
	
	\includegraphics{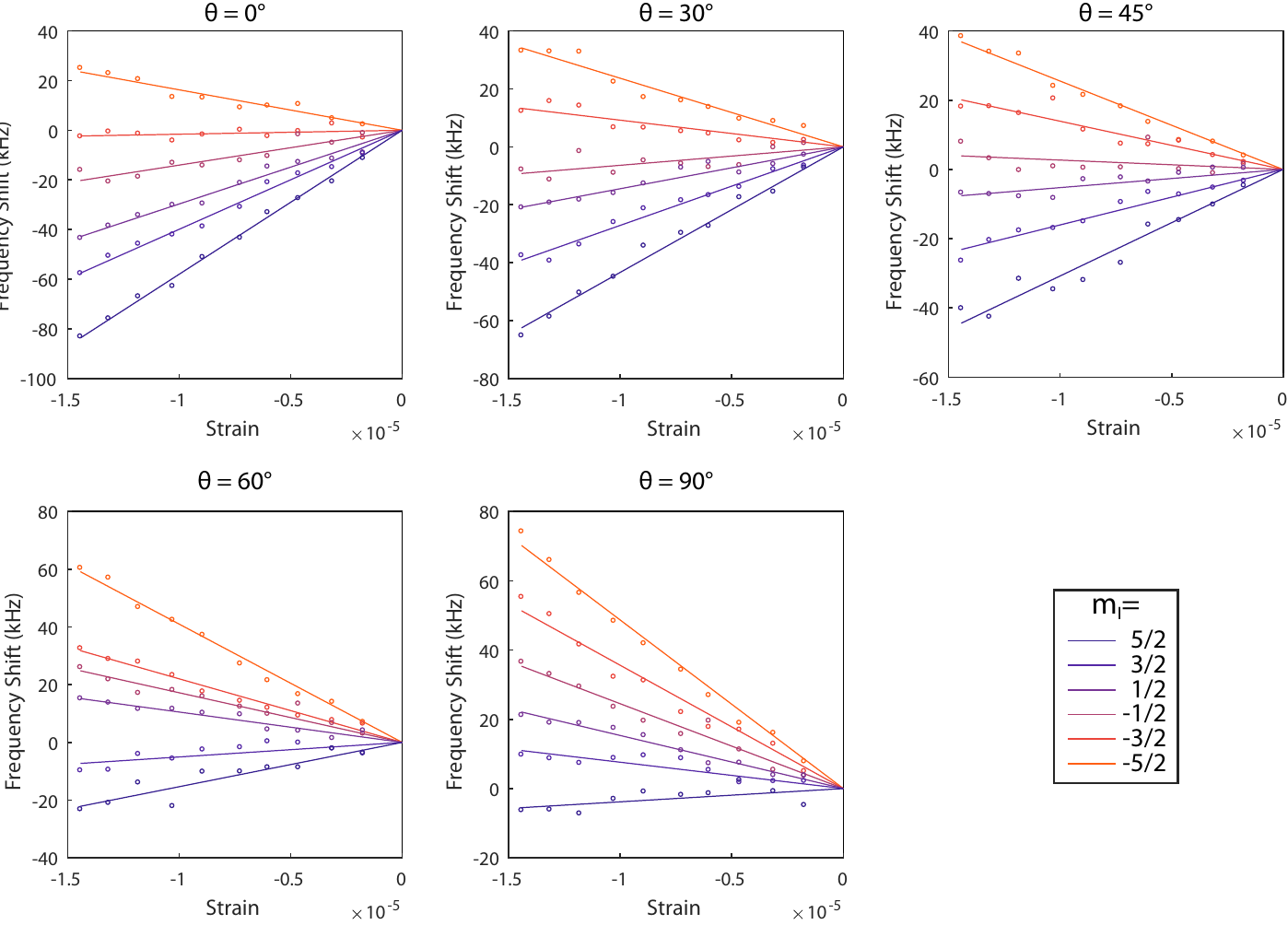}
	\caption{Full dataset for Si:Sb showing frequency shifts for all six ESR transitions as a function of strain $\varepsilon_{11}$ and $\theta$. Linear fits are shown for each transition. \label{fig:SiSb_full}}
\end{figure}

\begin{figure}[h]
	
	\includegraphics{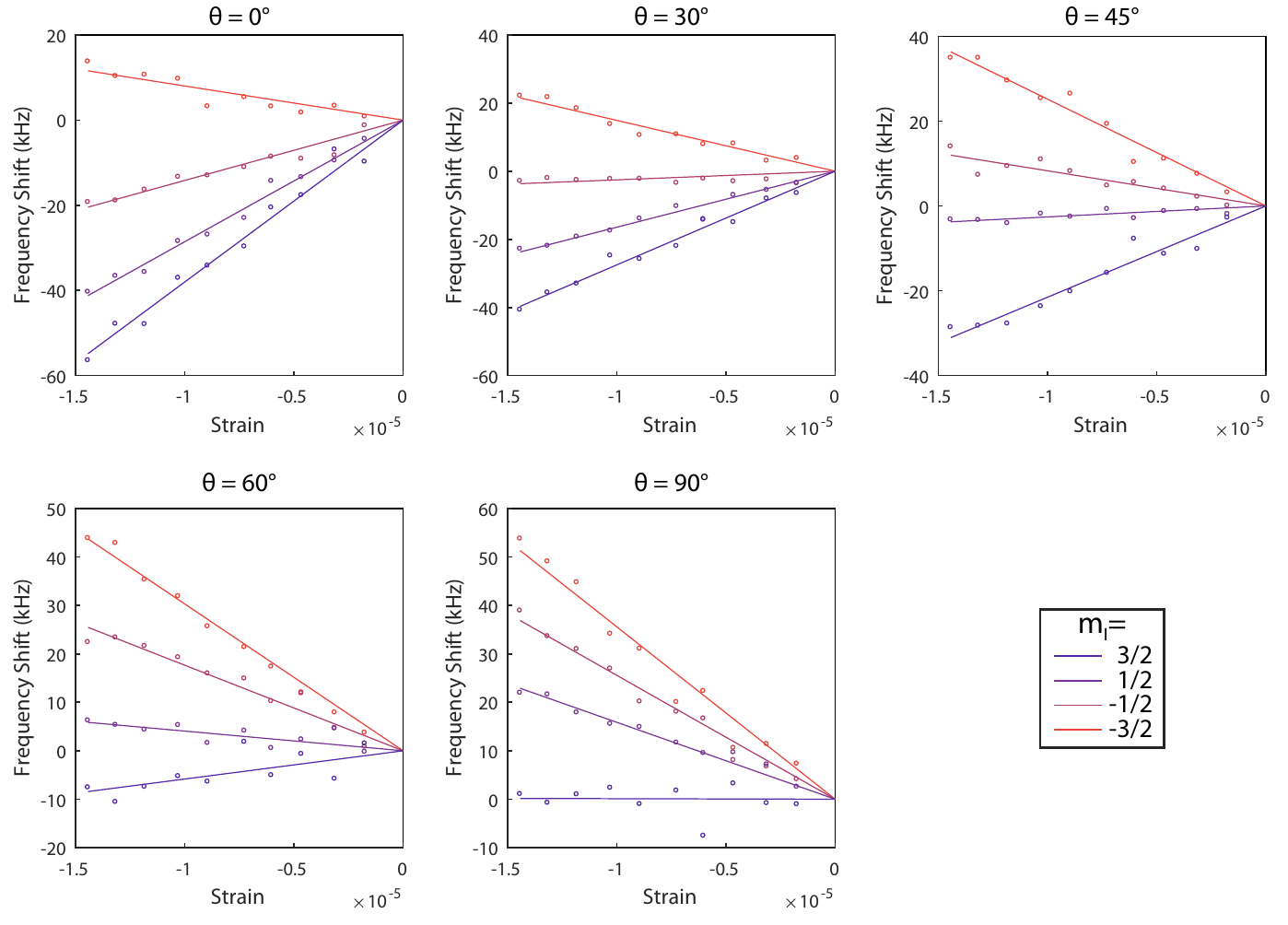}
	\caption{Full dataset for Si:As showing frequency shifts for all four ESR transitions as a function of strain $\varepsilon_{11}$ and $\theta$. Linear fits are shown for each transition. \label{fig:SiAs_full}}
\end{figure}

\begin{figure}[h]
	
	\includegraphics{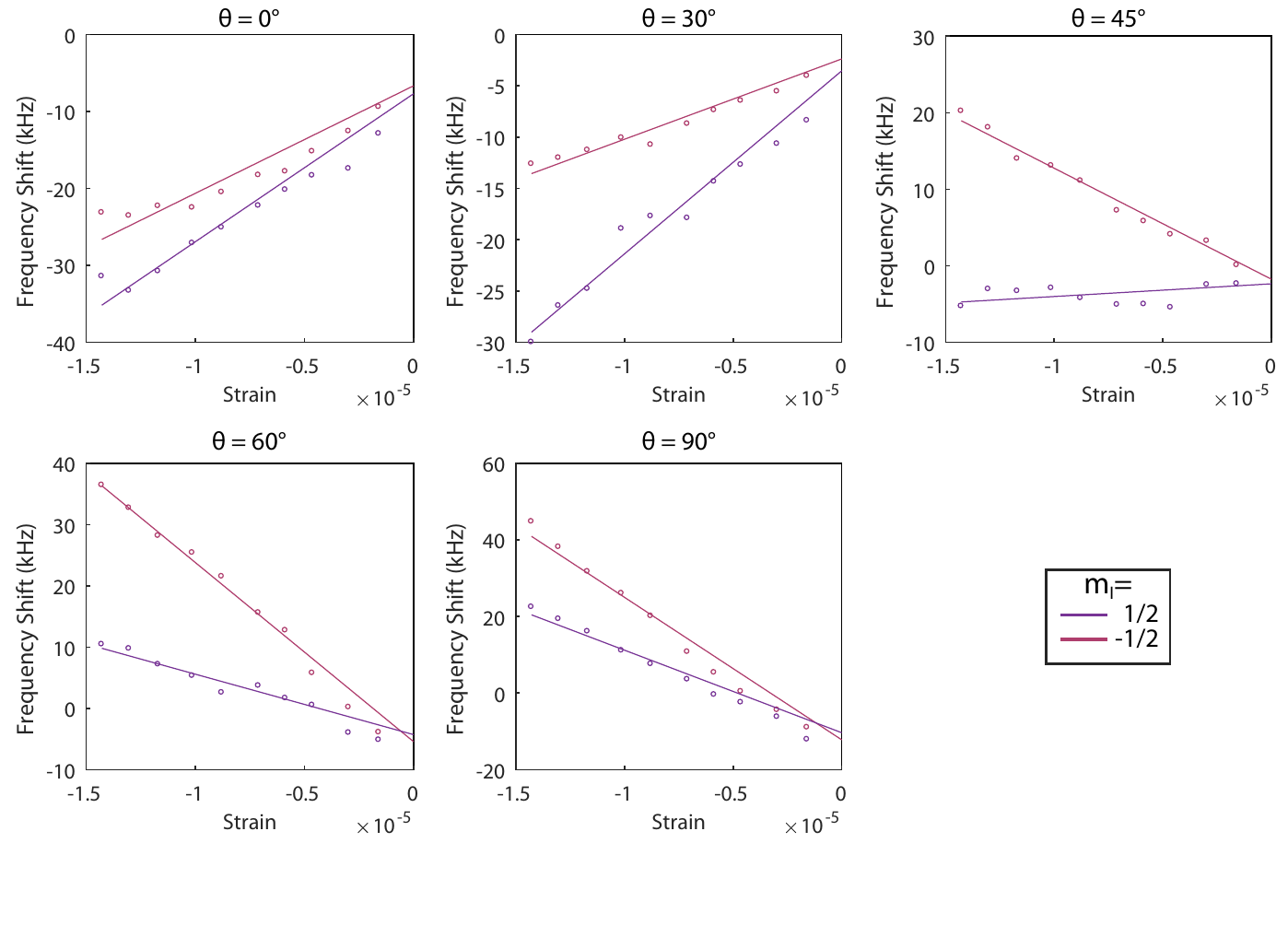}
	\caption{Full dataset for Si:P showing frequency shifts for both ESR transitions as a function of strain $\varepsilon_{11}$ and $\theta$. Linear fits are shown for each transition. \label{fig:SiP_full}}
\end{figure}

\bibliographystyle{apsrev4-1}

\begin{thebibliography}{72}%
	\makeatletter
	\providecommand \@ifxundefined [1]{%
		\@ifx{#1\undefined}
	}%
	\providecommand \@ifnum [1]{%
		\ifnum #1\expandafter \@firstoftwo
		\else \expandafter \@secondoftwo
		\fi
	}%
	\providecommand \@ifx [1]{%
		\ifx #1\expandafter \@firstoftwo
		\else \expandafter \@secondoftwo
		\fi
	}%
	\providecommand \natexlab [1]{#1}%
	\providecommand \enquote  [1]{``#1''}%
	\providecommand \bibnamefont  [1]{#1}%
	\providecommand \bibfnamefont [1]{#1}%
	\providecommand \citenamefont [1]{#1}%
	\providecommand \href@noop [0]{\@secondoftwo}%
	\providecommand \href [0]{\begingroup \@sanitize@url \@href}%
	\providecommand \@href[1]{\@@startlink{#1}\@@href}%
	\providecommand \@@href[1]{\endgroup#1\@@endlink}%
	\providecommand \@sanitize@url [0]{\catcode `\\12\catcode `\$12\catcode
		`\&12\catcode `\#12\catcode `\^12\catcode `\_12\catcode `\%12\relax}%
	\providecommand \@@startlink[1]{}%
	\providecommand \@@endlink[0]{}%
	\providecommand \url  [0]{\begingroup\@sanitize@url \@url }%
	\providecommand \@url [1]{\endgroup\@href {#1}{\urlprefix }}%
	\providecommand \urlprefix  [0]{URL }%
	\providecommand \Eprint [0]{\href }%
	\providecommand \doibase [0]{http://dx.doi.org/}%
	\providecommand \selectlanguage [0]{\@gobble}%
	\providecommand \bibinfo  [0]{\@secondoftwo}%
	\providecommand \bibfield  [0]{\@secondoftwo}%
	\providecommand \translation [1]{[#1]}%
	\providecommand \BibitemOpen [0]{}%
	\providecommand \bibitemStop [0]{}%
	\providecommand \bibitemNoStop [0]{.\EOS\space}%
	\providecommand \EOS [0]{\spacefactor3000\relax}%
	\providecommand \BibitemShut  [1]{\csname bibitem#1\endcsname}%
	\let\auto@bib@innerbib\@empty
	\bibitem [{\citenamefont {Tyryshkin}\ \emph {et~al.}(2011)\citenamefont
		{Tyryshkin}, \citenamefont {Tojo}, \citenamefont {Morton}, \citenamefont
		{Riemann}, \citenamefont {Abrosimov}, \citenamefont {Becker}, \citenamefont
		{Pohl}, \citenamefont {Schenkel}, \citenamefont {Thewalt},\ and\
		\citenamefont {Itoh}}]{Tyryshkin11}%
	\BibitemOpen
	\bibfield  {author} {\bibinfo {author} {\bibfnamefont {A.~M.}\ \bibnamefont
			{Tyryshkin}}, \bibinfo {author} {\bibfnamefont {S.}~\bibnamefont {Tojo}},
		\bibinfo {author} {\bibfnamefont {J.~J.~L.}\ \bibnamefont {Morton}}, \bibinfo
		{author} {\bibfnamefont {H.}~\bibnamefont {Riemann}}, \bibinfo {author}
		{\bibfnamefont {N.~V.}\ \bibnamefont {Abrosimov}}, \bibinfo {author}
		{\bibfnamefont {P.}~\bibnamefont {Becker}}, \bibinfo {author} {\bibfnamefont
			{H.-J.}\ \bibnamefont {Pohl}}, \bibinfo {author} {\bibfnamefont
			{T.}~\bibnamefont {Schenkel}}, \bibinfo {author} {\bibfnamefont {M.~L.~W.}\
			\bibnamefont {Thewalt}}, \ and\ \bibinfo {author} {\bibfnamefont {K.~M.}\
			\bibnamefont {Itoh}},\ }\href {\doibase 10.1038/nmat3182} {\bibfield
		{journal} {\bibinfo  {journal} {Nature Materials}\ }\textbf {\bibinfo
			{volume} {11}},\ \bibinfo {pages} {143} (\bibinfo {year} {2011})}\BibitemShut
	{NoStop}%
	\bibitem [{\citenamefont {Saeedi}\ \emph {et~al.}(2013)\citenamefont {Saeedi},
		\citenamefont {Simmons}, \citenamefont {Salvail}, \citenamefont {Dluhy},
		\citenamefont {Riemann}, \citenamefont {Abrosimov}, \citenamefont {Becker},
		\citenamefont {Pohl}, \citenamefont {Morton},\ and\ \citenamefont
		{Thewalt}}]{Saeedi13}%
	\BibitemOpen
	\bibfield  {author} {\bibinfo {author} {\bibfnamefont {K.}~\bibnamefont
			{Saeedi}}, \bibinfo {author} {\bibfnamefont {S.}~\bibnamefont {Simmons}},
		\bibinfo {author} {\bibfnamefont {J.~Z.}\ \bibnamefont {Salvail}}, \bibinfo
		{author} {\bibfnamefont {P.}~\bibnamefont {Dluhy}}, \bibinfo {author}
		{\bibfnamefont {H.}~\bibnamefont {Riemann}}, \bibinfo {author} {\bibfnamefont
			{N.~V.}\ \bibnamefont {Abrosimov}}, \bibinfo {author} {\bibfnamefont
			{P.}~\bibnamefont {Becker}}, \bibinfo {author} {\bibfnamefont {H.-J.}\
			\bibnamefont {Pohl}}, \bibinfo {author} {\bibfnamefont {J.~J.~L.}\
			\bibnamefont {Morton}}, \ and\ \bibinfo {author} {\bibfnamefont {M.~L.~W.}\
			\bibnamefont {Thewalt}},\ }\href {\doibase 10.1126/science.1239584}
	{\bibfield  {journal} {\bibinfo  {journal} {Science}\ }\textbf {\bibinfo
			{volume} {342}},\ \bibinfo {pages} {830} (\bibinfo {year}
		{2013})}\BibitemShut {NoStop}%
	\bibitem [{\citenamefont {Pla}\ \emph {et~al.}(2013)\citenamefont {Pla},
		\citenamefont {Tan}, \citenamefont {Dehollain}, \citenamefont {Lim},
		\citenamefont {Morton}, \citenamefont {Zwanenburg}, \citenamefont {Jamieson},
		\citenamefont {Dzurak},\ and\ \citenamefont {Morello}}]{Pla13}%
	\BibitemOpen
	\bibfield  {author} {\bibinfo {author} {\bibfnamefont {J.~J.}\ \bibnamefont
			{Pla}}, \bibinfo {author} {\bibfnamefont {K.~Y.}\ \bibnamefont {Tan}},
		\bibinfo {author} {\bibfnamefont {J.~P.}\ \bibnamefont {Dehollain}}, \bibinfo
		{author} {\bibfnamefont {W.~H.}\ \bibnamefont {Lim}}, \bibinfo {author}
		{\bibfnamefont {J.~J.~L.}\ \bibnamefont {Morton}}, \bibinfo {author}
		{\bibfnamefont {F.~A.}\ \bibnamefont {Zwanenburg}}, \bibinfo {author}
		{\bibfnamefont {D.~N.}\ \bibnamefont {Jamieson}}, \bibinfo {author}
		{\bibfnamefont {A.~S.}\ \bibnamefont {Dzurak}}, \ and\ \bibinfo {author}
		{\bibfnamefont {A.}~\bibnamefont {Morello}},\ }\href {\doibase
		10.1038/nature12011} {\bibfield  {journal} {\bibinfo  {journal} {Nature}\
		}\textbf {\bibinfo {volume} {496}},\ \bibinfo {pages} {334} (\bibinfo {year}
		{2013})}\BibitemShut {NoStop}%
	\bibitem [{\citenamefont {Muhonen}\ \emph {et~al.}(2015)\citenamefont
		{Muhonen}, \citenamefont {Laucht}, \citenamefont {Simmons}, \citenamefont
		{Dehollain}, \citenamefont {Kalra}, \citenamefont {Hudson}, \citenamefont
		{Freer}, \citenamefont {Itoh}, \citenamefont {Jamieson},\ and\ \citenamefont
		{Mccallum}}]{Muhonen15}%
	\BibitemOpen
	\bibfield  {author} {\bibinfo {author} {\bibfnamefont {J.~T.}\ \bibnamefont
			{Muhonen}}, \bibinfo {author} {\bibfnamefont {A.}~\bibnamefont {Laucht}},
		\bibinfo {author} {\bibfnamefont {S.}~\bibnamefont {Simmons}}, \bibinfo
		{author} {\bibfnamefont {J.~P.}\ \bibnamefont {Dehollain}}, \bibinfo {author}
		{\bibfnamefont {R.}~\bibnamefont {Kalra}}, \bibinfo {author} {\bibfnamefont
			{F.~E.}\ \bibnamefont {Hudson}}, \bibinfo {author} {\bibfnamefont
			{S.}~\bibnamefont {Freer}}, \bibinfo {author} {\bibfnamefont {K.~M.}\
			\bibnamefont {Itoh}}, \bibinfo {author} {\bibfnamefont {D.~N.}\ \bibnamefont
			{Jamieson}}, \ and\ \bibinfo {author} {\bibfnamefont {J.~C.}\ \bibnamefont
			{Mccallum}},\ }\href {\doibase 10.1088/0953-8984/27/15/154205} {\bibfield
		{journal} {\bibinfo  {journal} {Journal of Physics: Condensed Matter}\
		}\textbf {\bibinfo {volume} {27}},\ \bibinfo {pages} {154205} (\bibinfo
		{year} {2015})}\BibitemShut {NoStop}%
	\bibitem [{\citenamefont {Kane}(1998)}]{Kane98}%
	\BibitemOpen
	\bibfield  {author} {\bibinfo {author} {\bibfnamefont {B.~E.}\ \bibnamefont
			{Kane}},\ }\href {\doibase 10.1038/nature30156} {\bibfield  {journal}
		{\bibinfo  {journal} {Nature}\ }\textbf {\bibinfo {volume} {393}},\ \bibinfo
		{pages} {133} (\bibinfo {year} {1998})}\BibitemShut {NoStop}%
	\bibitem [{\citenamefont {O'Brien}\ \emph {et~al.}(2001)\citenamefont
		{O'Brien}, \citenamefont {Schofield}, \citenamefont {Simmons}, \citenamefont
		{Clark}, \citenamefont {Dzurak}, \citenamefont {Curson}, \citenamefont
		{Kane}, \citenamefont {McAlpine}, \citenamefont {Hawley},\ and\ \citenamefont
		{Brown}}]{Obrien01}%
	\BibitemOpen
	\bibfield  {author} {\bibinfo {author} {\bibfnamefont {J.~L.}\ \bibnamefont
			{O'Brien}}, \bibinfo {author} {\bibfnamefont {S.~R.}\ \bibnamefont
			{Schofield}}, \bibinfo {author} {\bibfnamefont {M.~Y.}\ \bibnamefont
			{Simmons}}, \bibinfo {author} {\bibfnamefont {R.~G.}\ \bibnamefont {Clark}},
		\bibinfo {author} {\bibfnamefont {A.~S.}\ \bibnamefont {Dzurak}}, \bibinfo
		{author} {\bibfnamefont {N.~J.}\ \bibnamefont {Curson}}, \bibinfo {author}
		{\bibfnamefont {B.~E.}\ \bibnamefont {Kane}}, \bibinfo {author}
		{\bibfnamefont {N.~S.}\ \bibnamefont {McAlpine}}, \bibinfo {author}
		{\bibfnamefont {M.~E.}\ \bibnamefont {Hawley}}, \ and\ \bibinfo {author}
		{\bibfnamefont {G.~W.}\ \bibnamefont {Brown}},\ }\href {\doibase
		10.1103/physrevb.64.161401} {\bibfield  {journal} {\bibinfo  {journal}
			{Physical Review B}\ }\textbf {\bibinfo {volume} {64}},\ \bibinfo {pages}
		{161401} (\bibinfo {year} {2001})}\BibitemShut {NoStop}%
	\bibitem [{\citenamefont {Tyryshkin}\ \emph {et~al.}(2003)\citenamefont
		{Tyryshkin}, \citenamefont {Lyon}, \citenamefont {Astashkin},\ and\
		\citenamefont {Raitsimring}}]{Tyryshkin03}%
	\BibitemOpen
	\bibfield  {author} {\bibinfo {author} {\bibfnamefont {A.~M.}\ \bibnamefont
			{Tyryshkin}}, \bibinfo {author} {\bibfnamefont {S.~A.}\ \bibnamefont {Lyon}},
		\bibinfo {author} {\bibfnamefont {A.~V.}\ \bibnamefont {Astashkin}}, \ and\
		\bibinfo {author} {\bibfnamefont {A.~M.}\ \bibnamefont {Raitsimring}},\
	}\href {\doibase 10.1103/physrevb.68.193207} {\bibfield  {journal} {\bibinfo
			{journal} {Physical Review B}\ }\textbf {\bibinfo {volume} {68}},\ \bibinfo
		{pages} {193207} (\bibinfo {year} {2003})}\BibitemShut {NoStop}%
	\bibitem [{\citenamefont {Stegner}\ \emph {et~al.}(2006)\citenamefont
		{Stegner}, \citenamefont {Boehme}, \citenamefont {Huebl}, \citenamefont
		{Stutzmann}, \citenamefont {Lips},\ and\ \citenamefont {Brandt}}]{Stegner06}%
	\BibitemOpen
	\bibfield  {author} {\bibinfo {author} {\bibfnamefont {A.~R.}\ \bibnamefont
			{Stegner}}, \bibinfo {author} {\bibfnamefont {C.}~\bibnamefont {Boehme}},
		\bibinfo {author} {\bibfnamefont {H.}~\bibnamefont {Huebl}}, \bibinfo
		{author} {\bibfnamefont {M.}~\bibnamefont {Stutzmann}}, \bibinfo {author}
		{\bibfnamefont {K.}~\bibnamefont {Lips}}, \ and\ \bibinfo {author}
		{\bibfnamefont {M.~S.}\ \bibnamefont {Brandt}},\ }\href {\doibase
		10.1038/nphys465} {\bibfield  {journal} {\bibinfo  {journal} {Nature
				Physics}\ }\textbf {\bibinfo {volume} {2}},\ \bibinfo {pages} {835} (\bibinfo
		{year} {2006})}\BibitemShut {NoStop}%
	\bibitem [{\citenamefont {Morton}\ \emph {et~al.}(2008)\citenamefont {Morton},
		\citenamefont {Tyryshkin}, \citenamefont {Brown}, \citenamefont {Shankar},
		\citenamefont {Lovett}, \citenamefont {Ardavan}, \citenamefont {Schenkel},
		\citenamefont {Haller}, \citenamefont {Ager},\ and\ \citenamefont
		{Lyon}}]{Morton08}%
	\BibitemOpen
	\bibfield  {author} {\bibinfo {author} {\bibfnamefont {J.~J.~L.}\
			\bibnamefont {Morton}}, \bibinfo {author} {\bibfnamefont {A.~M.}\
			\bibnamefont {Tyryshkin}}, \bibinfo {author} {\bibfnamefont {R.~M.}\
			\bibnamefont {Brown}}, \bibinfo {author} {\bibfnamefont {S.}~\bibnamefont
			{Shankar}}, \bibinfo {author} {\bibfnamefont {B.~W.}\ \bibnamefont {Lovett}},
		\bibinfo {author} {\bibfnamefont {A.}~\bibnamefont {Ardavan}}, \bibinfo
		{author} {\bibfnamefont {T.}~\bibnamefont {Schenkel}}, \bibinfo {author}
		{\bibfnamefont {E.~E.}\ \bibnamefont {Haller}}, \bibinfo {author}
		{\bibfnamefont {J.~W.}\ \bibnamefont {Ager}}, \ and\ \bibinfo {author}
		{\bibfnamefont {S.~A.}\ \bibnamefont {Lyon}},\ }\href {\doibase
		10.1038/nature07295} {\bibfield  {journal} {\bibinfo  {journal} {Nature}\
		}\textbf {\bibinfo {volume} {455}},\ \bibinfo {pages} {1085} (\bibinfo {year}
		{2008})}\BibitemShut {NoStop}%
	\bibitem [{\citenamefont {Simmons}\ \emph {et~al.}(2011)\citenamefont
		{Simmons}, \citenamefont {Brown}, \citenamefont {Riemann}, \citenamefont
		{Abrosimov}, \citenamefont {Becker}, \citenamefont {Pohl}, \citenamefont
		{Thewalt}, \citenamefont {Itoh},\ and\ \citenamefont {Morton}}]{Simmons11}%
	\BibitemOpen
	\bibfield  {author} {\bibinfo {author} {\bibfnamefont {S.}~\bibnamefont
			{Simmons}}, \bibinfo {author} {\bibfnamefont {R.~M.}\ \bibnamefont {Brown}},
		\bibinfo {author} {\bibfnamefont {H.}~\bibnamefont {Riemann}}, \bibinfo
		{author} {\bibfnamefont {N.~V.}\ \bibnamefont {Abrosimov}}, \bibinfo {author}
		{\bibfnamefont {P.}~\bibnamefont {Becker}}, \bibinfo {author} {\bibfnamefont
			{H.-J.}\ \bibnamefont {Pohl}}, \bibinfo {author} {\bibfnamefont {M.~L.~W.}\
			\bibnamefont {Thewalt}}, \bibinfo {author} {\bibfnamefont {K.~M.}\
			\bibnamefont {Itoh}}, \ and\ \bibinfo {author} {\bibfnamefont {J.~J.~L.}\
			\bibnamefont {Morton}},\ }\href {\doibase 10.1038/nature09696} {\bibfield
		{journal} {\bibinfo  {journal} {Nature}\ }\textbf {\bibinfo {volume} {470}},\
		\bibinfo {pages} {69} (\bibinfo {year} {2011})}\BibitemShut {NoStop}%
	\bibitem [{\citenamefont {Fuechsle}\ \emph {et~al.}(2012)\citenamefont
		{Fuechsle}, \citenamefont {Miwa}, \citenamefont {Mahapatra}, \citenamefont
		{Ryu}, \citenamefont {Lee}, \citenamefont {Warschkow}, \citenamefont
		{Hollenberg}, \citenamefont {Klimeck},\ and\ \citenamefont
		{Simmons}}]{Fuechsle12}%
	\BibitemOpen
	\bibfield  {author} {\bibinfo {author} {\bibfnamefont {M.}~\bibnamefont
			{Fuechsle}}, \bibinfo {author} {\bibfnamefont {J.~A.}\ \bibnamefont {Miwa}},
		\bibinfo {author} {\bibfnamefont {S.}~\bibnamefont {Mahapatra}}, \bibinfo
		{author} {\bibfnamefont {H.}~\bibnamefont {Ryu}}, \bibinfo {author}
		{\bibfnamefont {S.}~\bibnamefont {Lee}}, \bibinfo {author} {\bibfnamefont
			{O.}~\bibnamefont {Warschkow}}, \bibinfo {author} {\bibfnamefont {L.~C.~L.}\
			\bibnamefont {Hollenberg}}, \bibinfo {author} {\bibfnamefont
			{G.}~\bibnamefont {Klimeck}}, \ and\ \bibinfo {author} {\bibfnamefont
			{M.~Y.}\ \bibnamefont {Simmons}},\ }\href {\doibase 10.1038/nnano.2012.21}
	{\bibfield  {journal} {\bibinfo  {journal} {Nature Nanotechnology}\ }\textbf
		{\bibinfo {volume} {7}},\ \bibinfo {pages} {242} (\bibinfo {year}
		{2012})}\BibitemShut {NoStop}%
	\bibitem [{\citenamefont {B\"{u}ch}\ \emph {et~al.}(2013)\citenamefont
		{B\"{u}ch}, \citenamefont {Mahapatra}, \citenamefont {Rahman}, \citenamefont
		{Morello},\ and\ \citenamefont {Simmons}}]{Buch13}%
	\BibitemOpen
	\bibfield  {author} {\bibinfo {author} {\bibfnamefont {H.}~\bibnamefont
			{B\"{u}ch}}, \bibinfo {author} {\bibfnamefont {S.}~\bibnamefont {Mahapatra}},
		\bibinfo {author} {\bibfnamefont {R.}~\bibnamefont {Rahman}}, \bibinfo
		{author} {\bibfnamefont {A.}~\bibnamefont {Morello}}, \ and\ \bibinfo
		{author} {\bibfnamefont {M.~Y.}\ \bibnamefont {Simmons}},\ }\href {\doibase
		10.1038/ncomms3017} {\bibfield  {journal} {\bibinfo  {journal} {Nature
				Communications}\ }\textbf {\bibinfo {volume} {4}} (\bibinfo {year} {2013}),\
		10.1038/ncomms3017}\BibitemShut {NoStop}%
	\bibitem [{\citenamefont {Franke}\ \emph {et~al.}(2015)\citenamefont {Franke},
		\citenamefont {Hrubesch}, \citenamefont {K\"{u}nzl}, \citenamefont {Becker},
		\citenamefont {Itoh}, \citenamefont {Stutzmann}, \citenamefont {Hoehne},
		\citenamefont {Dreher},\ and\ \citenamefont {Brandt}}]{Franke15}%
	\BibitemOpen
	\bibfield  {author} {\bibinfo {author} {\bibfnamefont {D.~P.}\ \bibnamefont
			{Franke}}, \bibinfo {author} {\bibfnamefont {F.~M.}\ \bibnamefont
			{Hrubesch}}, \bibinfo {author} {\bibfnamefont {M.}~\bibnamefont {K\"{u}nzl}},
		\bibinfo {author} {\bibfnamefont {H.-W.}\ \bibnamefont {Becker}}, \bibinfo
		{author} {\bibfnamefont {K.~M.}\ \bibnamefont {Itoh}}, \bibinfo {author}
		{\bibfnamefont {M.}~\bibnamefont {Stutzmann}}, \bibinfo {author}
		{\bibfnamefont {F.}~\bibnamefont {Hoehne}}, \bibinfo {author} {\bibfnamefont
			{L.}~\bibnamefont {Dreher}}, \ and\ \bibinfo {author} {\bibfnamefont {M.~S.}\
			\bibnamefont {Brandt}},\ }\href {\doibase 10.1103/physrevlett.115.057601}
	{\bibfield  {journal} {\bibinfo  {journal} {Physical Review Letters}\
		}\textbf {\bibinfo {volume} {115}},\ \bibinfo {pages} {057601} (\bibinfo
		{year} {2015})}\BibitemShut {NoStop}%
	\bibitem [{\citenamefont {Franke}\ \emph {et~al.}(2016)\citenamefont {Franke},
		\citenamefont {Pfl\"{u}ger}, \citenamefont {Mortemousque}, \citenamefont
		{Itoh},\ and\ \citenamefont {Brandt}}]{Franke16}%
	\BibitemOpen
	\bibfield  {author} {\bibinfo {author} {\bibfnamefont {D.~P.}\ \bibnamefont
			{Franke}}, \bibinfo {author} {\bibfnamefont {M.~P.~D.}\ \bibnamefont
			{Pfl\"{u}ger}}, \bibinfo {author} {\bibfnamefont {P.-A.}\ \bibnamefont
			{Mortemousque}}, \bibinfo {author} {\bibfnamefont {K.~M.}\ \bibnamefont
			{Itoh}}, \ and\ \bibinfo {author} {\bibfnamefont {M.~S.}\ \bibnamefont
			{Brandt}},\ }\href {\doibase 10.1103/physrevb.93.161303} {\bibfield
		{journal} {\bibinfo  {journal} {Physical Review B}\ }\textbf {\bibinfo
			{volume} {93}},\ \bibinfo {pages} {161303} (\bibinfo {year}
		{2016})}\BibitemShut {NoStop}%
	\bibitem [{\citenamefont {Wolfowicz}\ \emph {et~al.}(2014)\citenamefont
		{Wolfowicz}, \citenamefont {Urdampilleta}, \citenamefont {Thewalt},
		\citenamefont {Riemann}, \citenamefont {Abrosimov}, \citenamefont {Becker},
		\citenamefont {Pohl},\ and\ \citenamefont {Morton}}]{Wolfowicz14}%
	\BibitemOpen
	\bibfield  {author} {\bibinfo {author} {\bibfnamefont {G.}~\bibnamefont
			{Wolfowicz}}, \bibinfo {author} {\bibfnamefont {M.}~\bibnamefont
			{Urdampilleta}}, \bibinfo {author} {\bibfnamefont {M.~L.~W.}\ \bibnamefont
			{Thewalt}}, \bibinfo {author} {\bibfnamefont {H.}~\bibnamefont {Riemann}},
		\bibinfo {author} {\bibfnamefont {N.~V.}\ \bibnamefont {Abrosimov}}, \bibinfo
		{author} {\bibfnamefont {P.}~\bibnamefont {Becker}}, \bibinfo {author}
		{\bibfnamefont {H.-J.}\ \bibnamefont {Pohl}}, \ and\ \bibinfo {author}
		{\bibfnamefont {J.~J.~L.}\ \bibnamefont {Morton}},\ }\href {\doibase
		10.1103/physrevlett.113.157601} {\bibfield  {journal} {\bibinfo  {journal}
			{Physical Review Letters}\ }\textbf {\bibinfo {volume} {113}},\ \bibinfo
		{pages} {157601} (\bibinfo {year} {2014})}\BibitemShut {NoStop}%
	\bibitem [{\citenamefont {Singh}\ \emph {et~al.}(2016)\citenamefont {Singh},
		\citenamefont {Pacheco}, \citenamefont {Perry}, \citenamefont {Garratt},
		\citenamefont {Eyck}, \citenamefont {Bishop}, \citenamefont {Wendt},
		\citenamefont {Manginell}, \citenamefont {Dominguez},\ and\ \citenamefont
		{Pluym}}]{Singh16}%
	\BibitemOpen
	\bibfield  {author} {\bibinfo {author} {\bibfnamefont {M.}~\bibnamefont
			{Singh}}, \bibinfo {author} {\bibfnamefont {J.~L.}\ \bibnamefont {Pacheco}},
		\bibinfo {author} {\bibfnamefont {D.}~\bibnamefont {Perry}}, \bibinfo
		{author} {\bibfnamefont {E.}~\bibnamefont {Garratt}}, \bibinfo {author}
		{\bibfnamefont {G.~T.}\ \bibnamefont {Eyck}}, \bibinfo {author}
		{\bibfnamefont {N.~C.}\ \bibnamefont {Bishop}}, \bibinfo {author}
		{\bibfnamefont {J.~R.}\ \bibnamefont {Wendt}}, \bibinfo {author}
		{\bibfnamefont {R.~P.}\ \bibnamefont {Manginell}}, \bibinfo {author}
		{\bibfnamefont {J.}~\bibnamefont {Dominguez}}, \ and\ \bibinfo {author}
		{\bibfnamefont {T.}~\bibnamefont {Pluym}},\ }\href {\doibase
		10.1063/1.4940421} {\bibfield  {journal} {\bibinfo  {journal} {Applied
				Physics Letters}\ }\textbf {\bibinfo {volume} {108}},\ \bibinfo {pages}
		{062101} (\bibinfo {year} {2016})}\BibitemShut {NoStop}%
	\bibitem [{\citenamefont {Morley}\ \emph {et~al.}(2010)\citenamefont {Morley},
		\citenamefont {Warner}, \citenamefont {Stoneham}, \citenamefont {Greenland},
		\citenamefont {Tol}, \citenamefont {Kay},\ and\ \citenamefont
		{Aeppli}}]{Morley10}%
	\BibitemOpen
	\bibfield  {author} {\bibinfo {author} {\bibfnamefont {G.~W.}\ \bibnamefont
			{Morley}}, \bibinfo {author} {\bibfnamefont {M.}~\bibnamefont {Warner}},
		\bibinfo {author} {\bibfnamefont {A.~M.}\ \bibnamefont {Stoneham}}, \bibinfo
		{author} {\bibfnamefont {P.~T.}\ \bibnamefont {Greenland}}, \bibinfo {author}
		{\bibfnamefont {J.~V.}\ \bibnamefont {Tol}}, \bibinfo {author} {\bibfnamefont
			{C.~W.~M.}\ \bibnamefont {Kay}}, \ and\ \bibinfo {author} {\bibfnamefont
			{G.}~\bibnamefont {Aeppli}},\ }\href {\doibase 10.1038/nmat2828} {\bibfield
		{journal} {\bibinfo  {journal} {Nature Materials}\ }\textbf {\bibinfo
			{volume} {9}},\ \bibinfo {pages} {725} (\bibinfo {year} {2010})}\BibitemShut
	{NoStop}%
	\bibitem [{\citenamefont {Mortemousque}\ \emph {et~al.}(2014)\citenamefont
		{Mortemousque}, \citenamefont {Berger}, \citenamefont {Sekiguchi},
		\citenamefont {Culan}, \citenamefont {Elliman},\ and\ \citenamefont
		{Itoh}}]{Mortemousque14}%
	\BibitemOpen
	\bibfield  {author} {\bibinfo {author} {\bibfnamefont {P.~A.}\ \bibnamefont
			{Mortemousque}}, \bibinfo {author} {\bibfnamefont {S.}~\bibnamefont
			{Berger}}, \bibinfo {author} {\bibfnamefont {T.}~\bibnamefont {Sekiguchi}},
		\bibinfo {author} {\bibfnamefont {C.}~\bibnamefont {Culan}}, \bibinfo
		{author} {\bibfnamefont {R.~G.}\ \bibnamefont {Elliman}}, \ and\ \bibinfo
		{author} {\bibfnamefont {K.~M.}\ \bibnamefont {Itoh}},\ }\href {\doibase
		10.1103/physrevb.89.155202} {\bibfield  {journal} {\bibinfo  {journal}
			{Physical Review B}\ }\textbf {\bibinfo {volume} {89}},\ \bibinfo {pages}
		{155202} (\bibinfo {year} {2014})}\BibitemShut {NoStop}%
	\bibitem [{\citenamefont {Saeedi}\ \emph {et~al.}(2015)\citenamefont {Saeedi},
		\citenamefont {Szech}, \citenamefont {Dluhy}, \citenamefont {Salvail},
		\citenamefont {Morse}, \citenamefont {Riemann}, \citenamefont {Abrosimov},
		\citenamefont {N\"{o}tzel}, \citenamefont {Litvinenko}, \citenamefont
		{Murdin},\ and\ \citenamefont {Thewalt}}]{Saeedi15}%
	\BibitemOpen
	\bibfield  {author} {\bibinfo {author} {\bibfnamefont {K.}~\bibnamefont
			{Saeedi}}, \bibinfo {author} {\bibfnamefont {M.}~\bibnamefont {Szech}},
		\bibinfo {author} {\bibfnamefont {P.}~\bibnamefont {Dluhy}}, \bibinfo
		{author} {\bibfnamefont {J.}~\bibnamefont {Salvail}}, \bibinfo {author}
		{\bibfnamefont {K.}~\bibnamefont {Morse}}, \bibinfo {author} {\bibfnamefont
			{H.}~\bibnamefont {Riemann}}, \bibinfo {author} {\bibfnamefont
			{N.}~\bibnamefont {Abrosimov}}, \bibinfo {author} {\bibfnamefont
			{N.}~\bibnamefont {N\"{o}tzel}}, \bibinfo {author} {\bibfnamefont
			{K.}~\bibnamefont {Litvinenko}}, \bibinfo {author} {\bibfnamefont
			{B.}~\bibnamefont {Murdin}}, \ and\ \bibinfo {author} {\bibfnamefont
			{M.}~\bibnamefont {Thewalt}},\ }\href {\doibase 10.1038/srep10493} {\bibfield
		{journal} {\bibinfo  {journal} {Scientific Reports}\ }\textbf {\bibinfo
			{volume} {5}} (\bibinfo {year} {2015}),\ 10.1038/srep10493}\BibitemShut
	{NoStop}%
	\bibitem [{\citenamefont {Wolfowicz}\ \emph {et~al.}(2013)\citenamefont
		{Wolfowicz}, \citenamefont {Tyryshkin}, \citenamefont {George}, \citenamefont
		{Riemann}, \citenamefont {Abrosimov}, \citenamefont {Becker}, \citenamefont
		{Pohl}, \citenamefont {Thewalt}, \citenamefont {Lyon},\ and\ \citenamefont
		{Morton}}]{Wolfowicz13}%
	\BibitemOpen
	\bibfield  {author} {\bibinfo {author} {\bibfnamefont {G.}~\bibnamefont
			{Wolfowicz}}, \bibinfo {author} {\bibfnamefont {A.~M.}\ \bibnamefont
			{Tyryshkin}}, \bibinfo {author} {\bibfnamefont {R.~E.}\ \bibnamefont
			{George}}, \bibinfo {author} {\bibfnamefont {H.}~\bibnamefont {Riemann}},
		\bibinfo {author} {\bibfnamefont {N.~V.}\ \bibnamefont {Abrosimov}}, \bibinfo
		{author} {\bibfnamefont {P.}~\bibnamefont {Becker}}, \bibinfo {author}
		{\bibfnamefont {H.-J.}\ \bibnamefont {Pohl}}, \bibinfo {author}
		{\bibfnamefont {M.~L.~W.}\ \bibnamefont {Thewalt}}, \bibinfo {author}
		{\bibfnamefont {S.~A.}\ \bibnamefont {Lyon}}, \ and\ \bibinfo {author}
		{\bibfnamefont {J.~J.~L.}\ \bibnamefont {Morton}},\ }\href {\doibase
		10.1038/nnano.2013.218} {\bibfield  {journal} {\bibinfo  {journal} {Nature
				Nanotechnology}\ }\textbf {\bibinfo {volume} {8}},\ \bibinfo {pages} {881}
		(\bibinfo {year} {2013})}\BibitemShut {NoStop}%
	\bibitem [{\citenamefont {Morello}\ \emph {et~al.}(2010)\citenamefont
		{Morello}, \citenamefont {Pla}, \citenamefont {Zwanenburg}, \citenamefont
		{Chan}, \citenamefont {Tan}, \citenamefont {Huebl}, \citenamefont
		{M\"{o}tt\"{o}nen}, \citenamefont {Nugroho}, \citenamefont {Yang},
		\citenamefont {van Donkelaar}, \citenamefont {Alves}, \citenamefont
		{Jamieson}, \citenamefont {Escott}, \citenamefont {Hollenberg}, \citenamefont
		{Clark},\ and\ \citenamefont {Dzurak}}]{Morello10}%
	\BibitemOpen
	\bibfield  {author} {\bibinfo {author} {\bibfnamefont {A.}~\bibnamefont
			{Morello}}, \bibinfo {author} {\bibfnamefont {J.~J.}\ \bibnamefont {Pla}},
		\bibinfo {author} {\bibfnamefont {F.~A.}\ \bibnamefont {Zwanenburg}},
		\bibinfo {author} {\bibfnamefont {K.~W.}\ \bibnamefont {Chan}}, \bibinfo
		{author} {\bibfnamefont {K.~Y.}\ \bibnamefont {Tan}}, \bibinfo {author}
		{\bibfnamefont {H.}~\bibnamefont {Huebl}}, \bibinfo {author} {\bibfnamefont
			{M.}~\bibnamefont {M\"{o}tt\"{o}nen}}, \bibinfo {author} {\bibfnamefont
			{C.~D.}\ \bibnamefont {Nugroho}}, \bibinfo {author} {\bibfnamefont
			{C.}~\bibnamefont {Yang}}, \bibinfo {author} {\bibfnamefont {J.~A.}\
			\bibnamefont {van Donkelaar}}, \bibinfo {author} {\bibfnamefont {A.~D.}\
			\bibnamefont {Alves}}, \bibinfo {author} {\bibfnamefont {D.~N.}\ \bibnamefont
			{Jamieson}}, \bibinfo {author} {\bibfnamefont {C.~C.}\ \bibnamefont
			{Escott}}, \bibinfo {author} {\bibfnamefont {L.~C.~L.}\ \bibnamefont
			{Hollenberg}}, \bibinfo {author} {\bibfnamefont {R.~G.}\ \bibnamefont
			{Clark}}, \ and\ \bibinfo {author} {\bibfnamefont {A.~S.}\ \bibnamefont
			{Dzurak}},\ }\href {\doibase 10.1038/nature09392} {\bibfield  {journal}
		{\bibinfo  {journal} {Nature}\ }\textbf {\bibinfo {volume} {467}},\ \bibinfo
		{pages} {687} (\bibinfo {year} {2010})}\BibitemShut {NoStop}%
	\bibitem [{\citenamefont {Pla}\ \emph {et~al.}(2012)\citenamefont {Pla},
		\citenamefont {Tan}, \citenamefont {Dehollain}, \citenamefont {Lim},
		\citenamefont {Morton}, \citenamefont {Jamieson}, \citenamefont {Dzurak},\
		and\ \citenamefont {Morello}}]{Pla12}%
	\BibitemOpen
	\bibfield  {author} {\bibinfo {author} {\bibfnamefont {J.~J.}\ \bibnamefont
			{Pla}}, \bibinfo {author} {\bibfnamefont {K.~Y.}\ \bibnamefont {Tan}},
		\bibinfo {author} {\bibfnamefont {J.~P.}\ \bibnamefont {Dehollain}}, \bibinfo
		{author} {\bibfnamefont {W.~H.}\ \bibnamefont {Lim}}, \bibinfo {author}
		{\bibfnamefont {J.~J.~L.}\ \bibnamefont {Morton}}, \bibinfo {author}
		{\bibfnamefont {D.~N.}\ \bibnamefont {Jamieson}}, \bibinfo {author}
		{\bibfnamefont {A.~S.}\ \bibnamefont {Dzurak}}, \ and\ \bibinfo {author}
		{\bibfnamefont {A.}~\bibnamefont {Morello}},\ }\href {\doibase
		10.1038/nature11449} {\bibfield  {journal} {\bibinfo  {journal} {Nature}\
		}\textbf {\bibinfo {volume} {489}},\ \bibinfo {pages} {541} (\bibinfo {year}
		{2012})}\BibitemShut {NoStop}%
	\bibitem [{\citenamefont {Bienfait}\ \emph {et~al.}(2016)\citenamefont
		{Bienfait}, \citenamefont {Pla}, \citenamefont {Kubo}, \citenamefont {Zhou},
		\citenamefont {Stern}, \citenamefont {Lo}, \citenamefont {Weis},
		\citenamefont {Schenkel}, \citenamefont {Vion}, \citenamefont {Esteve},
		\citenamefont {Morton},\ and\ \citenamefont {Bertet}}]{Bienfait16}%
	\BibitemOpen
	\bibfield  {author} {\bibinfo {author} {\bibfnamefont {A.}~\bibnamefont
			{Bienfait}}, \bibinfo {author} {\bibfnamefont {J.~J.}\ \bibnamefont {Pla}},
		\bibinfo {author} {\bibfnamefont {Y.}~\bibnamefont {Kubo}}, \bibinfo {author}
		{\bibfnamefont {X.}~\bibnamefont {Zhou}}, \bibinfo {author} {\bibfnamefont
			{M.}~\bibnamefont {Stern}}, \bibinfo {author} {\bibfnamefont {C.~C.}\
			\bibnamefont {Lo}}, \bibinfo {author} {\bibfnamefont {C.~D.}\ \bibnamefont
			{Weis}}, \bibinfo {author} {\bibfnamefont {T.}~\bibnamefont {Schenkel}},
		\bibinfo {author} {\bibfnamefont {D.}~\bibnamefont {Vion}}, \bibinfo {author}
		{\bibfnamefont {D.}~\bibnamefont {Esteve}}, \bibinfo {author} {\bibfnamefont
			{J.}~\bibnamefont {Morton}}, \ and\ \bibinfo {author} {\bibfnamefont
			{P.}~\bibnamefont {Bertet}},\ }\href {\doibase 10.1038/nature16944}
	{\bibfield  {journal} {\bibinfo  {journal} {Nature}\ }\textbf {\bibinfo
			{volume} {531}},\ \bibinfo {pages} {74} (\bibinfo {year} {2016})}\BibitemShut
	{NoStop}%
	\bibitem [{\citenamefont {Eichler}\ \emph {et~al.}(2017)\citenamefont
		{Eichler}, \citenamefont {Sigillito}, \citenamefont {Lyon},\ and\
		\citenamefont {Petta}}]{Eichler17}%
	\BibitemOpen
	\bibfield  {author} {\bibinfo {author} {\bibfnamefont {C.}~\bibnamefont
			{Eichler}}, \bibinfo {author} {\bibfnamefont {A.~J.}\ \bibnamefont
			{Sigillito}}, \bibinfo {author} {\bibfnamefont {S.~A.}\ \bibnamefont {Lyon}},
		\ and\ \bibinfo {author} {\bibfnamefont {J.~R.}\ \bibnamefont {Petta}},\
	}\href {\doibase 10.1103/physrevlett.118.037701} {\bibfield  {journal}
		{\bibinfo  {journal} {Physical Review Letters}\ }\textbf {\bibinfo {volume}
			{118}},\ \bibinfo {pages} {037701} (\bibinfo {year} {2017})}\BibitemShut
	{NoStop}%
	\bibitem [{\citenamefont {Zollitsch}\ \emph {et~al.}(2015)\citenamefont
		{Zollitsch}, \citenamefont {Mueller}, \citenamefont {Franke}, \citenamefont
		{Goennenwein}, \citenamefont {Brandt}, \citenamefont {Gross},\ and\
		\citenamefont {Huebl}}]{Zollitsch15}%
	\BibitemOpen
	\bibfield  {author} {\bibinfo {author} {\bibfnamefont {C.~W.}\ \bibnamefont
			{Zollitsch}}, \bibinfo {author} {\bibfnamefont {K.}~\bibnamefont {Mueller}},
		\bibinfo {author} {\bibfnamefont {D.~P.}\ \bibnamefont {Franke}}, \bibinfo
		{author} {\bibfnamefont {S.~T.~B.}\ \bibnamefont {Goennenwein}}, \bibinfo
		{author} {\bibfnamefont {M.~S.}\ \bibnamefont {Brandt}}, \bibinfo {author}
		{\bibfnamefont {R.}~\bibnamefont {Gross}}, \ and\ \bibinfo {author}
		{\bibfnamefont {H.}~\bibnamefont {Huebl}},\ }\href {\doibase
		10.1063/1.4932658} {\bibfield  {journal} {\bibinfo  {journal} {Applied
				Physics Letters}\ }\textbf {\bibinfo {volume} {107}},\ \bibinfo {pages}
		{142105} (\bibinfo {year} {2015})}\BibitemShut {NoStop}%
	\bibitem [{\citenamefont {Angus}\ \emph {et~al.}(2007)\citenamefont {Angus},
		\citenamefont {Ferguson}, \citenamefont {Dzurak},\ and\ \citenamefont
		{Clark}}]{Angus07}%
	\BibitemOpen
	\bibfield  {author} {\bibinfo {author} {\bibfnamefont {S.~J.}\ \bibnamefont
			{Angus}}, \bibinfo {author} {\bibfnamefont {A.~J.}\ \bibnamefont {Ferguson}},
		\bibinfo {author} {\bibfnamefont {A.~S.}\ \bibnamefont {Dzurak}}, \ and\
		\bibinfo {author} {\bibfnamefont {R.~G.}\ \bibnamefont {Clark}},\ }\href
	{\doibase 10.1021/nl070949k} {\bibfield  {journal} {\bibinfo  {journal} {Nano
				Letters}\ }\textbf {\bibinfo {volume} {7}},\ \bibinfo {pages} {2051}
		(\bibinfo {year} {2007})}\BibitemShut {NoStop}%
	\bibitem [{\citenamefont {Bruno}\ \emph {et~al.}(2015)\citenamefont {Bruno},
		\citenamefont {Lange}, \citenamefont {Asaad}, \citenamefont {Enden},
		\citenamefont {Langford},\ and\ \citenamefont {Dicarlo}}]{Bruno15}%
	\BibitemOpen
	\bibfield  {author} {\bibinfo {author} {\bibfnamefont {A.}~\bibnamefont
			{Bruno}}, \bibinfo {author} {\bibfnamefont {G.~D.}\ \bibnamefont {Lange}},
		\bibinfo {author} {\bibfnamefont {S.}~\bibnamefont {Asaad}}, \bibinfo
		{author} {\bibfnamefont {K.~L. V.~D.}\ \bibnamefont {Enden}}, \bibinfo
		{author} {\bibfnamefont {N.~K.}\ \bibnamefont {Langford}}, \ and\ \bibinfo
		{author} {\bibfnamefont {L.}~\bibnamefont {Dicarlo}},\ }\href {\doibase
		10.1063/1.4919761} {\bibfield  {journal} {\bibinfo  {journal} {Applied
				Physics Letters}\ }\textbf {\bibinfo {volume} {106}},\ \bibinfo {pages}
		{182601} (\bibinfo {year} {2015})}\BibitemShut {NoStop}%
	\bibitem [{\citenamefont {Lyon}\ \emph {et~al.}(1977)\citenamefont {Lyon},
		\citenamefont {Salinger}, \citenamefont {Swenson},\ and\ \citenamefont
		{White}}]{Lyon77}%
	\BibitemOpen
	\bibfield  {author} {\bibinfo {author} {\bibfnamefont {K.~G.}\ \bibnamefont
			{Lyon}}, \bibinfo {author} {\bibfnamefont {G.~L.}\ \bibnamefont {Salinger}},
		\bibinfo {author} {\bibfnamefont {C.~A.}\ \bibnamefont {Swenson}}, \ and\
		\bibinfo {author} {\bibfnamefont {G.~K.}\ \bibnamefont {White}},\ }\href
	{\doibase 10.1063/1.323747} {\bibfield  {journal} {\bibinfo  {journal}
			{Journal of Applied Physics}\ }\textbf {\bibinfo {volume} {48}},\ \bibinfo
		{pages} {865} (\bibinfo {year} {1977})}\BibitemShut {NoStop}%
	\bibitem [{\citenamefont {Swenson}(1983)}]{Swenson83}%
	\BibitemOpen
	\bibfield  {author} {\bibinfo {author} {\bibfnamefont {C.~A.}\ \bibnamefont
			{Swenson}},\ }\href {\doibase 10.1063/1.555681} {\bibfield  {journal}
		{\bibinfo  {journal} {Journal of Physical and Chemical Reference Data}\
		}\textbf {\bibinfo {volume} {12}},\ \bibinfo {pages} {179} (\bibinfo {year}
		{1983})}\BibitemShut {NoStop}%
	\bibitem [{\citenamefont {Roberts}(1982)}]{Roberts82}%
	\BibitemOpen
	\bibfield  {author} {\bibinfo {author} {\bibfnamefont {R.~B.}\ \bibnamefont
			{Roberts}},\ }\href {\doibase 10.1088/0022-3727/15/9/004} {\bibfield
		{journal} {\bibinfo  {journal} {Journal of Physics D: Applied Physics}\
		}\textbf {\bibinfo {volume} {15}} (\bibinfo {year} {1982}),\
		10.1088/0022-3727/15/9/004}\BibitemShut {NoStop}%
	\bibitem [{NIS(1991)}]{NIST91}%
	\BibitemOpen
	\href@noop {} {\emph {\bibinfo {title} {NIST Standard Reference Material 739,
				Certificate}}} (\bibinfo {year} {1991})\BibitemShut {NoStop}%
	\bibitem [{\citenamefont {Nix}\ and\ \citenamefont
		{Macnair}(1941)}]{NixMacnair41}%
	\BibitemOpen
	\bibfield  {author} {\bibinfo {author} {\bibfnamefont {F.~C.}\ \bibnamefont
			{Nix}}\ and\ \bibinfo {author} {\bibfnamefont {D.}~\bibnamefont {Macnair}},\
	}\href {\doibase 10.1103/physrev.60.597} {\bibfield  {journal} {\bibinfo
			{journal} {Physical Review}\ }\textbf {\bibinfo {volume} {60}},\ \bibinfo
		{pages} {597} (\bibinfo {year} {1941})}\BibitemShut {NoStop}%
	\bibitem [{\citenamefont {Morello}\ \emph {et~al.}(2009)\citenamefont
		{Morello}, \citenamefont {Escott}, \citenamefont {Huebl}, \citenamefont
		{Willems~van Beveren}, \citenamefont {Hollenberg}, \citenamefont {Jamieson},
		\citenamefont {Dzurak},\ and\ \citenamefont {Clark}}]{Morello09}%
	\BibitemOpen
	\bibfield  {author} {\bibinfo {author} {\bibfnamefont {A.}~\bibnamefont
			{Morello}}, \bibinfo {author} {\bibfnamefont {C.~C.}\ \bibnamefont {Escott}},
		\bibinfo {author} {\bibfnamefont {H.}~\bibnamefont {Huebl}}, \bibinfo
		{author} {\bibfnamefont {L.~H.}\ \bibnamefont {Willems~van Beveren}},
		\bibinfo {author} {\bibfnamefont {L.~C.~L.}\ \bibnamefont {Hollenberg}},
		\bibinfo {author} {\bibfnamefont {D.~N.}\ \bibnamefont {Jamieson}}, \bibinfo
		{author} {\bibfnamefont {A.~S.}\ \bibnamefont {Dzurak}}, \ and\ \bibinfo
		{author} {\bibfnamefont {R.~G.}\ \bibnamefont {Clark}},\ }\href {\doibase
		10.1103/physrevb.80.081307} {\bibfield  {journal} {\bibinfo  {journal}
			{Physical Review B}\ }\textbf {\bibinfo {volume} {80}},\ \bibinfo {pages}
		{081307} (\bibinfo {year} {2009})}\BibitemShut {NoStop}%
	\bibitem [{\citenamefont {Pla}\ \emph {et~al.}(2016)\citenamefont {Pla},
		\citenamefont {Bienfait}, \citenamefont {Pica}, \citenamefont {Mansir},
		\citenamefont {Mohiyaddin}, \citenamefont {Morello}, \citenamefont
		{Schenkel}, \citenamefont {Lovett}, \citenamefont {Morton},\ and\
		\citenamefont {Bertet}}]{1608.07346}%
	\BibitemOpen
	\bibfield  {author} {\bibinfo {author} {\bibfnamefont {J.~J.}\ \bibnamefont
			{Pla}}, \bibinfo {author} {\bibfnamefont {A.}~\bibnamefont {Bienfait}},
		\bibinfo {author} {\bibfnamefont {G.}~\bibnamefont {Pica}}, \bibinfo {author}
		{\bibfnamefont {J.}~\bibnamefont {Mansir}}, \bibinfo {author} {\bibfnamefont
			{F.~A.}\ \bibnamefont {Mohiyaddin}}, \bibinfo {author} {\bibfnamefont
			{A.}~\bibnamefont {Morello}}, \bibinfo {author} {\bibfnamefont
			{T.}~\bibnamefont {Schenkel}}, \bibinfo {author} {\bibfnamefont {B.~W.}\
			\bibnamefont {Lovett}}, \bibinfo {author} {\bibfnamefont {J.~J.~L.}\
			\bibnamefont {Morton}}, \ and\ \bibinfo {author} {\bibfnamefont
			{P.}~\bibnamefont {Bertet}},\ }\href@noop {} {\enquote {\bibinfo {title}
			{Strain-induced nuclear quadrupole splittings in silicon devices},}\ }
	(\bibinfo {year} {2016}),\ \Eprint {http://arxiv.org/abs/arXiv:1608.07346}
	{arXiv:1608.07346} \BibitemShut {NoStop}%
	\bibitem [{\citenamefont {Bardeen}\ and\ \citenamefont
		{Shockley}(1950)}]{BardeenShockley50}%
	\BibitemOpen
	\bibfield  {author} {\bibinfo {author} {\bibfnamefont {J.}~\bibnamefont
			{Bardeen}}\ and\ \bibinfo {author} {\bibfnamefont {W.}~\bibnamefont
			{Shockley}},\ }\href {\doibase 10.1103/physrev.80.72} {\bibfield  {journal}
		{\bibinfo  {journal} {Physical Review}\ }\textbf {\bibinfo {volume} {80}},\
		\bibinfo {pages} {72} (\bibinfo {year} {1950})}\BibitemShut {NoStop}%
	\bibitem [{\citenamefont {Herring}\ and\ \citenamefont
		{Vogt}(1957)}]{HerringVogt57}%
	\BibitemOpen
	\bibfield  {author} {\bibinfo {author} {\bibfnamefont {C.}~\bibnamefont
			{Herring}}\ and\ \bibinfo {author} {\bibfnamefont {E.}~\bibnamefont {Vogt}},\
	}\href {\doibase 10.1103/physrev.105.1933} {\bibfield  {journal} {\bibinfo
			{journal} {Physical Review}\ }\textbf {\bibinfo {volume} {105}},\ \bibinfo
		{pages} {1933} (\bibinfo {year} {1957})}\BibitemShut {NoStop}%
	\bibitem [{\citenamefont {Thorbeck}\ and\ \citenamefont
		{Zimmerman}(2015)}]{Thorbeck15}%
	\BibitemOpen
	\bibfield  {author} {\bibinfo {author} {\bibfnamefont {T.}~\bibnamefont
			{Thorbeck}}\ and\ \bibinfo {author} {\bibfnamefont {N.~M.}\ \bibnamefont
			{Zimmerman}},\ }\href {\doibase 10.1063/1.4928320} {\bibfield  {journal}
		{\bibinfo  {journal} {AIP Advances}\ }\textbf {\bibinfo {volume} {5}},\
		\bibinfo {pages} {087107} (\bibinfo {year} {2015})}\BibitemShut {NoStop}%
	\bibitem [{\citenamefont {Veldhorst}\ \emph {et~al.}(2015)\citenamefont
		{Veldhorst}, \citenamefont {Yang}, \citenamefont {Hwang}, \citenamefont
		{Huang}, \citenamefont {Dehollain}, \citenamefont {Muhonen}, \citenamefont
		{Simmons}, \citenamefont {Laucht}, \citenamefont {Hudson},\ and\
		\citenamefont {Itoh}}]{Veldhorst15}%
	\BibitemOpen
	\bibfield  {author} {\bibinfo {author} {\bibfnamefont {M.}~\bibnamefont
			{Veldhorst}}, \bibinfo {author} {\bibfnamefont {C.~H.}\ \bibnamefont {Yang}},
		\bibinfo {author} {\bibfnamefont {J.~C.~C.}\ \bibnamefont {Hwang}}, \bibinfo
		{author} {\bibfnamefont {W.}~\bibnamefont {Huang}}, \bibinfo {author}
		{\bibfnamefont {J.~P.}\ \bibnamefont {Dehollain}}, \bibinfo {author}
		{\bibfnamefont {J.~T.}\ \bibnamefont {Muhonen}}, \bibinfo {author}
		{\bibfnamefont {S.}~\bibnamefont {Simmons}}, \bibinfo {author} {\bibfnamefont
			{A.}~\bibnamefont {Laucht}}, \bibinfo {author} {\bibfnamefont {F.~E.}\
			\bibnamefont {Hudson}}, \ and\ \bibinfo {author} {\bibfnamefont {K.~M.}\
			\bibnamefont {Itoh}},\ }\href {\doibase 10.1038/nature15263} {\bibfield
		{journal} {\bibinfo  {journal} {Nature}\ }\textbf {\bibinfo {volume} {526}},\
		\bibinfo {pages} {410} (\bibinfo {year} {2015})}\BibitemShut {NoStop}%
	\bibitem [{\citenamefont {Wilson}\ and\ \citenamefont
		{Feher}(1961{\natexlab{a}})}]{WilsonFeher61}%
	\BibitemOpen
	\bibfield  {author} {\bibinfo {author} {\bibfnamefont {D.~K.}\ \bibnamefont
			{Wilson}}\ and\ \bibinfo {author} {\bibfnamefont {G.}~\bibnamefont {Feher}},\
	}\href {\doibase 10.1103/physrev.124.1068} {\bibfield  {journal} {\bibinfo
			{journal} {Physical Review}\ }\textbf {\bibinfo {volume} {124}},\ \bibinfo
		{pages} {1068} (\bibinfo {year} {1961}{\natexlab{a}})}\BibitemShut {NoStop}%
	\bibitem [{\citenamefont {Huebl}\ \emph {et~al.}(2006)\citenamefont {Huebl},
		\citenamefont {Stegner}, \citenamefont {Stutzmann}, \citenamefont {Brandt},
		\citenamefont {Vogg}, \citenamefont {Bensch}, \citenamefont {Rauls},\ and\
		\citenamefont {Gerstmann}}]{Huebl06}%
	\BibitemOpen
	\bibfield  {author} {\bibinfo {author} {\bibfnamefont {H.}~\bibnamefont
			{Huebl}}, \bibinfo {author} {\bibfnamefont {A.~R.}\ \bibnamefont {Stegner}},
		\bibinfo {author} {\bibfnamefont {M.}~\bibnamefont {Stutzmann}}, \bibinfo
		{author} {\bibfnamefont {M.~S.}\ \bibnamefont {Brandt}}, \bibinfo {author}
		{\bibfnamefont {G.}~\bibnamefont {Vogg}}, \bibinfo {author} {\bibfnamefont
			{F.}~\bibnamefont {Bensch}}, \bibinfo {author} {\bibfnamefont
			{E.}~\bibnamefont {Rauls}}, \ and\ \bibinfo {author} {\bibfnamefont
			{U.}~\bibnamefont {Gerstmann}},\ }\href {\doibase
		10.1103/PhysRevLett.97.166402} {\bibfield  {journal} {\bibinfo  {journal}
			{Phys. Rev. Lett.}\ }\textbf {\bibinfo {volume} {97}},\ \bibinfo {pages}
		{166402} (\bibinfo {year} {2006})}\BibitemShut {NoStop}%
	\bibitem [{\citenamefont {Dreher}\ \emph {et~al.}(2011)\citenamefont {Dreher},
		\citenamefont {Hilker}, \citenamefont {Brandlmaier}, \citenamefont
		{Goennenwein}, \citenamefont {Huebl}, \citenamefont {Stutzmann},\ and\
		\citenamefont {Brandt}}]{Dreher11}%
	\BibitemOpen
	\bibfield  {author} {\bibinfo {author} {\bibfnamefont {L.}~\bibnamefont
			{Dreher}}, \bibinfo {author} {\bibfnamefont {T.~A.}\ \bibnamefont {Hilker}},
		\bibinfo {author} {\bibfnamefont {A.}~\bibnamefont {Brandlmaier}}, \bibinfo
		{author} {\bibfnamefont {S.~T.~B.}\ \bibnamefont {Goennenwein}}, \bibinfo
		{author} {\bibfnamefont {H.}~\bibnamefont {Huebl}}, \bibinfo {author}
		{\bibfnamefont {M.}~\bibnamefont {Stutzmann}}, \ and\ \bibinfo {author}
		{\bibfnamefont {M.~S.}\ \bibnamefont {Brandt}},\ }\href {\doibase
		10.1103/physrevlett.106.037601} {\bibfield  {journal} {\bibinfo  {journal}
			{Physical Review Letters}\ }\textbf {\bibinfo {volume} {106}},\ \bibinfo
		{pages} {037601} (\bibinfo {year} {2011})}\BibitemShut {NoStop}%
	\bibitem [{\citenamefont {Pines}\ \emph {et~al.}(1957)\citenamefont {Pines},
		\citenamefont {Bardeen},\ and\ \citenamefont {Slichter}}]{PBS57_main}%
	\BibitemOpen
	\bibfield  {author} {\bibinfo {author} {\bibfnamefont {D.}~\bibnamefont
			{Pines}}, \bibinfo {author} {\bibfnamefont {J.}~\bibnamefont {Bardeen}}, \
		and\ \bibinfo {author} {\bibfnamefont {C.~P.}\ \bibnamefont {Slichter}},\
	}\href {\doibase 10.1103/PhysRev.106.489} {\bibfield  {journal} {\bibinfo
			{journal} {Phys. Rev.}\ }\textbf {\bibinfo {volume} {106}},\ \bibinfo {pages}
		{489} (\bibinfo {year} {1957})}\BibitemShut {NoStop}%
	\bibitem [{\citenamefont {Schweiger}\ and\ \citenamefont
		{Jeschke}(2005)}]{Schweiger05_main}%
	\BibitemOpen
	\bibfield  {author} {\bibinfo {author} {\bibfnamefont {A.}~\bibnamefont
			{Schweiger}}\ and\ \bibinfo {author} {\bibfnamefont {G.}~\bibnamefont
			{Jeschke}},\ }\href@noop {} {\emph {\bibinfo {title} {Principles of pulse
				electron paramagnetic resonance}}}\ (\bibinfo  {publisher} {Oxford University
		Press},\ \bibinfo {year} {2005})\BibitemShut {NoStop}%
	\bibitem [{sup()}]{suppmat}%
	\BibitemOpen
	\href@noop {} {\bibinfo  {journal} {{See Supplemental Material for further
				details on experimental methods, tight-binding modeling, DFT calculations,
				modeling the g-factor anisotropy, and complete datasets. }}\ }\BibitemShut {NoStop}%
	\bibitem [{\citenamefont {Zhang}\ \emph {et~al.}(2014)\citenamefont {Zhang},
		\citenamefont {Barrett}, \citenamefont {Cloetens}, \citenamefont {Detlefs},\
		and\ \citenamefont {Rio}}]{Zhang14}%
	\BibitemOpen
	\bibfield  {journal} {  }\bibfield  {author} {\bibinfo {author} {\bibfnamefont
			{L.}~\bibnamefont {Zhang}}, \bibinfo {author} {\bibfnamefont
			{R.}~\bibnamefont {Barrett}}, \bibinfo {author} {\bibfnamefont
			{P.}~\bibnamefont {Cloetens}}, \bibinfo {author} {\bibfnamefont
			{C.}~\bibnamefont {Detlefs}}, \ and\ \bibinfo {author} {\bibfnamefont
			{M.~S.~D.}\ \bibnamefont {Rio}},\ }\href {\doibase 10.1107/s1600577514004962}
	{\bibfield  {journal} {\bibinfo  {journal} {Journal of Synchrotron
				Radiation}\ }\textbf {\bibinfo {volume} {21}},\ \bibinfo {pages} {507}
		(\bibinfo {year} {2014})}\BibitemShut {NoStop}%
	\bibitem [{\citenamefont {Ross}(2017)}]{Ross17}%
	\BibitemOpen
	\bibfield  {author} {\bibinfo {author} {\bibfnamefont {P.}~\bibnamefont
			{Ross}},\ }\href@noop {} {Ph.D. thesis},\ \bibinfo  {school} {University
		College London} (\bibinfo {year} {2017})\BibitemShut {NoStop}%
	\bibitem [{\citenamefont {Hahn}(1950)}]{Hahn50}%
	\BibitemOpen
	\bibfield  {author} {\bibinfo {author} {\bibfnamefont {E.~L.}\ \bibnamefont
			{Hahn}},\ }\href {\doibase 10.1103/physrev.80.580} {\bibfield  {journal}
		{\bibinfo  {journal} {Physical Review}\ }\textbf {\bibinfo {volume} {80}},\
		\bibinfo {pages} {580} (\bibinfo {year} {1950})}\BibitemShut {NoStop}%
	\bibitem [{\citenamefont {Kohn}\ and\ \citenamefont
		{Luttinger}(1955)}]{KohnLuttinger55}%
	\BibitemOpen
	\bibfield  {author} {\bibinfo {author} {\bibfnamefont {W.}~\bibnamefont
			{Kohn}}\ and\ \bibinfo {author} {\bibfnamefont {J.~M.}\ \bibnamefont
			{Luttinger}},\ }\href {\doibase 10.1103/physrev.98.915} {\bibfield  {journal}
		{\bibinfo  {journal} {Physical Review}\ }\textbf {\bibinfo {volume} {98}},\
		\bibinfo {pages} {915} (\bibinfo {year} {1955})}\BibitemShut {NoStop}%
	\bibitem [{\citenamefont {Luttinger}\ and\ \citenamefont
		{Kohn}(1955)}]{LuttingerKohn55}%
	\BibitemOpen
	\bibfield  {author} {\bibinfo {author} {\bibfnamefont {J.~M.}\ \bibnamefont
			{Luttinger}}\ and\ \bibinfo {author} {\bibfnamefont {W.}~\bibnamefont
			{Kohn}},\ }\href {\doibase 10.1103/physrev.97.869} {\bibfield  {journal}
		{\bibinfo  {journal} {Physical Review}\ }\textbf {\bibinfo {volume} {97}},\
		\bibinfo {pages} {869} (\bibinfo {year} {1955})}\BibitemShut {NoStop}%
	\bibitem [{\citenamefont {Pantelides}\ and\ \citenamefont
		{Sah}(1974)}]{Pantelides74}%
	\BibitemOpen
	\bibfield  {author} {\bibinfo {author} {\bibfnamefont {S.~T.}\ \bibnamefont
			{Pantelides}}\ and\ \bibinfo {author} {\bibfnamefont {C.~T.}\ \bibnamefont
			{Sah}},\ }\href {\doibase 10.1103/physrevb.10.621} {\bibfield  {journal}
		{\bibinfo  {journal} {Physical Review B}\ }\textbf {\bibinfo {volume} {10}},\
		\bibinfo {pages} {621} (\bibinfo {year} {1974})}\BibitemShut {NoStop}%
	\bibitem [{\citenamefont {Pica}\ \emph
		{et~al.}(2014{\natexlab{a}})\citenamefont {Pica}, \citenamefont {Wolfowicz},
		\citenamefont {Urdampilleta}, \citenamefont {Thewalt}, \citenamefont
		{Riemann}, \citenamefont {Abrosimov}, \citenamefont {Becker}, \citenamefont
		{Pohl}, \citenamefont {Morton}, \citenamefont {Bhatt}, \citenamefont {Lyon},\
		and\ \citenamefont {Lovett}}]{Pica14}%
	\BibitemOpen
	\bibfield  {author} {\bibinfo {author} {\bibfnamefont {G.}~\bibnamefont
			{Pica}}, \bibinfo {author} {\bibfnamefont {G.}~\bibnamefont {Wolfowicz}},
		\bibinfo {author} {\bibfnamefont {M.}~\bibnamefont {Urdampilleta}}, \bibinfo
		{author} {\bibfnamefont {M.~L.~W.}\ \bibnamefont {Thewalt}}, \bibinfo
		{author} {\bibfnamefont {H.}~\bibnamefont {Riemann}}, \bibinfo {author}
		{\bibfnamefont {N.~V.}\ \bibnamefont {Abrosimov}}, \bibinfo {author}
		{\bibfnamefont {P.}~\bibnamefont {Becker}}, \bibinfo {author} {\bibfnamefont
			{H.-J.}\ \bibnamefont {Pohl}}, \bibinfo {author} {\bibfnamefont {J.~J.~L.}\
			\bibnamefont {Morton}}, \bibinfo {author} {\bibfnamefont {R.~N.}\
			\bibnamefont {Bhatt}}, \bibinfo {author} {\bibfnamefont {S.~A.}\ \bibnamefont
			{Lyon}}, \ and\ \bibinfo {author} {\bibfnamefont {B.~W.}\ \bibnamefont
			{Lovett}},\ }\href@noop {} {\bibfield  {journal} {\bibinfo  {journal}
			{Physical Review B}\ }\textbf {\bibinfo {volume} {90}},\ \bibinfo {pages}
		{195204} (\bibinfo {year} {2014}{\natexlab{a}})}\BibitemShut {NoStop}%
	\bibitem [{\citenamefont {Pica}\ \emph
		{et~al.}(2014{\natexlab{b}})\citenamefont {Pica}, \citenamefont {Lovett},
		\citenamefont {Bhatt},\ and\ \citenamefont {Lyon}}]{Pica14_2}%
	\BibitemOpen
	\bibfield  {author} {\bibinfo {author} {\bibfnamefont {G.}~\bibnamefont
			{Pica}}, \bibinfo {author} {\bibfnamefont {B.~W.}\ \bibnamefont {Lovett}},
		\bibinfo {author} {\bibfnamefont {R.~N.}\ \bibnamefont {Bhatt}}, \ and\
		\bibinfo {author} {\bibfnamefont {S.~A.}\ \bibnamefont {Lyon}},\ }\href
	{\doibase 10.1103/physrevb.89.235306} {\bibfield  {journal} {\bibinfo
			{journal} {Physical Review B}\ }\textbf {\bibinfo {volume} {89}},\ \bibinfo
		{pages} {235306} (\bibinfo {year} {2014}{\natexlab{b}})}\BibitemShut
	{NoStop}%
\end{thebibliography}

\clearpage

\begin{thebibliography}{24}%
	\makeatletter
	\providecommand \@ifxundefined [1]{%
		\@ifx{#1\undefined}
	}%
	\providecommand \@ifnum [1]{%
		\ifnum #1\expandafter \@firstoftwo
		\else \expandafter \@secondoftwo
		\fi
	}%
	\providecommand \@ifx [1]{%
		\ifx #1\expandafter \@firstoftwo
		\else \expandafter \@secondoftwo
		\fi
	}%
	\providecommand \natexlab [1]{#1}%
	\providecommand \enquote  [1]{``#1''}%
	\providecommand \bibnamefont  [1]{#1}%
	\providecommand \bibfnamefont [1]{#1}%
	\providecommand \citenamefont [1]{#1}%
	\providecommand \href@noop [0]{\@secondoftwo}%
	\providecommand \href [0]{\begingroup \@sanitize@url \@href}%
	\providecommand \@href[1]{\@@startlink{#1}\@@href}%
	\providecommand \@@href[1]{\endgroup#1\@@endlink}%
	\providecommand \@sanitize@url [0]{\catcode `\\12\catcode `\$12\catcode
		`\&12\catcode `\#12\catcode `\^12\catcode `\_12\catcode `\%12\relax}%
	\providecommand \@@startlink[1]{}%
	\providecommand \@@endlink[0]{}%
	\providecommand \url  [0]{\begingroup\@sanitize@url \@url }%
	\providecommand \@url [1]{\endgroup\@href {#1}{\urlprefix }}%
	\providecommand \urlprefix  [0]{URL }%
	\providecommand \Eprint [0]{\href }%
	\providecommand \doibase [0]{http://dx.doi.org/}%
	\providecommand \selectlanguage [0]{\@gobble}%
	\providecommand \bibinfo  [0]{\@secondoftwo}%
	\providecommand \bibfield  [0]{\@secondoftwo}%
	\providecommand \translation [1]{[#1]}%
	\providecommand \BibitemOpen [0]{}%
	\providecommand \bibitemStop [0]{}%
	\providecommand \bibitemNoStop [0]{.\EOS\space}%
	\providecommand \EOS [0]{\spacefactor3000\relax}%
	\providecommand \BibitemShut  [1]{\csname bibitem#1\endcsname}%
	\let\auto@bib@innerbib\@empty
	\bibitem [{\citenamefont {Schweiger}\ and\ \citenamefont
		{Jeschke}(2005)}]{Schweiger05}%
	\BibitemOpen
	\bibfield  {author} {\bibinfo {author} {\bibfnamefont {A.}~\bibnamefont
			{Schweiger}}\ and\ \bibinfo {author} {\bibfnamefont {G.}~\bibnamefont
			{Jeschke}},\ }\href@noop {} {\emph {\bibinfo {title} {Principles of pulse
				electron paramagnetic resonance}}}\ (\bibinfo  {publisher} {Oxford University
		Press},\ \bibinfo {year} {2005})\BibitemShut {NoStop}%
	\bibitem [{\citenamefont {Niquet}\ \emph {et~al.}(2009)\citenamefont {Niquet},
		\citenamefont {Rideau}, \citenamefont {Tavernier}, \citenamefont {Jaouen},\
		and\ \citenamefont {Blase}}]{Niquet09}%
	\BibitemOpen
	\bibfield  {author} {\bibinfo {author} {\bibfnamefont {Y.~M.}\ \bibnamefont
			{Niquet}}, \bibinfo {author} {\bibfnamefont {D.}~\bibnamefont {Rideau}},
		\bibinfo {author} {\bibfnamefont {C.}~\bibnamefont {Tavernier}}, \bibinfo
		{author} {\bibfnamefont {H.}~\bibnamefont {Jaouen}}, \ and\ \bibinfo {author}
		{\bibfnamefont {X.}~\bibnamefont {Blase}},\ }\href@noop {} {\bibfield
		{journal} {\bibinfo  {journal} {Physical Review B}\ }\textbf {\bibinfo
			{volume} {79}},\ \bibinfo {pages} {245201} (\bibinfo {year}
		{2009})}\BibitemShut {NoStop}%
	\bibitem [{\citenamefont {Roche}\ \emph {et~al.}(2012)\citenamefont {Roche},
		\citenamefont {Dupont-Ferrier}, \citenamefont {Voisin}, \citenamefont
		{Cobian}, \citenamefont {Jehl}, \citenamefont {Wacquez}, \citenamefont
		{Vinet}, \citenamefont {Niquet},\ and\ \citenamefont {Sanquer}}]{Roche12}%
	\BibitemOpen
	\bibfield  {author} {\bibinfo {author} {\bibfnamefont {B.}~\bibnamefont
			{Roche}}, \bibinfo {author} {\bibfnamefont {E.}~\bibnamefont
			{Dupont-Ferrier}}, \bibinfo {author} {\bibfnamefont {B.}~\bibnamefont
			{Voisin}}, \bibinfo {author} {\bibfnamefont {M.}~\bibnamefont {Cobian}},
		\bibinfo {author} {\bibfnamefont {X.}~\bibnamefont {Jehl}}, \bibinfo {author}
		{\bibfnamefont {R.}~\bibnamefont {Wacquez}}, \bibinfo {author} {\bibfnamefont
			{M.}~\bibnamefont {Vinet}}, \bibinfo {author} {\bibfnamefont {Y.-M.}\
			\bibnamefont {Niquet}}, \ and\ \bibinfo {author} {\bibfnamefont
			{M.}~\bibnamefont {Sanquer}},\ }\href@noop {} {\bibfield  {journal} {\bibinfo
			{journal} {Physical Review Letters}\ }\textbf {\bibinfo {volume} {108}},\
		\bibinfo {pages} {206812} (\bibinfo {year} {2012})}\BibitemShut {NoStop}%
	\bibitem [{\citenamefont {Usman}\ \emph {et~al.}(2015)\citenamefont {Usman},
		\citenamefont {Rahman}, \citenamefont {Salfi}, \citenamefont {Bocquel},
		\citenamefont {Voisin}, \citenamefont {Rogge}, \citenamefont {Klimeck},\ and\
		\citenamefont {Hollenberg}}]{Usman15}%
	\BibitemOpen
	\bibfield  {author} {\bibinfo {author} {\bibfnamefont {M.}~\bibnamefont
			{Usman}}, \bibinfo {author} {\bibfnamefont {R.}~\bibnamefont {Rahman}},
		\bibinfo {author} {\bibfnamefont {J.}~\bibnamefont {Salfi}}, \bibinfo
		{author} {\bibfnamefont {J.}~\bibnamefont {Bocquel}}, \bibinfo {author}
		{\bibfnamefont {B.}~\bibnamefont {Voisin}}, \bibinfo {author} {\bibfnamefont
			{S.}~\bibnamefont {Rogge}}, \bibinfo {author} {\bibfnamefont
			{G.}~\bibnamefont {Klimeck}}, \ and\ \bibinfo {author} {\bibfnamefont
			{L.~L.~C.}\ \bibnamefont {Hollenberg}},\ }\href@noop {} {\bibfield  {journal}
		{\bibinfo  {journal} {Journal of Physics: Condensed Matter}\ }\textbf
		{\bibinfo {volume} {27}},\ \bibinfo {pages} {154207} (\bibinfo {year}
		{2015})}\BibitemShut {NoStop}%
	\bibitem [{\citenamefont {Nara}\ and\ \citenamefont {Morita}(1966)}]{Nara66}%
	\BibitemOpen
	\bibfield  {author} {\bibinfo {author} {\bibfnamefont {H.}~\bibnamefont
			{Nara}}\ and\ \bibinfo {author} {\bibfnamefont {A.}~\bibnamefont {Morita}},\
	}\href@noop {} {\bibfield  {journal} {\bibinfo  {journal} {Journal of the
				Physical Society of Japan}\ }\textbf {\bibinfo {volume} {21}},\ \bibinfo
		{pages} {1852} (\bibinfo {year} {1966})}\BibitemShut {NoStop}%
	\bibitem [{\citenamefont {Bernholc}\ and\ \citenamefont
		{Pantelides}(1977)}]{Bernholc77}%
	\BibitemOpen
	\bibfield  {author} {\bibinfo {author} {\bibfnamefont {J.}~\bibnamefont
			{Bernholc}}\ and\ \bibinfo {author} {\bibfnamefont {S.~T.}\ \bibnamefont
			{Pantelides}},\ }\href {\doibase 10.1103/PhysRevB.15.4935} {\bibfield
		{journal} {\bibinfo  {journal} {Physical Review B}\ }\textbf {\bibinfo
			{volume} {15}},\ \bibinfo {pages} {4935} (\bibinfo {year}
		{1977})}\BibitemShut {NoStop}%
	\bibitem [{\citenamefont {Krag}\ \emph {et~al.}(1970)\citenamefont {Krag},
		\citenamefont {Kleiner},\ and\ \citenamefont {Zieger}}]{Krag70}%
	\BibitemOpen
	\bibfield  {author} {\bibinfo {author} {\bibfnamefont {W.~E.}\ \bibnamefont
			{Krag}}, \bibinfo {author} {\bibfnamefont {W.~H.}\ \bibnamefont {Kleiner}}, \
		and\ \bibinfo {author} {\bibfnamefont {H.~J.}\ \bibnamefont {Zieger}},\ }in\
	\href@noop {} {\emph {\bibinfo {booktitle} {{Proceedings of the 10th
					International Conference on the Physics of Semiconductors (Cambridge,
					Massachusetts, 1970)}}}},\ \bibinfo {editor} {edited by\ \bibinfo {editor}
		{\bibfnamefont {S.~P.}\ \bibnamefont {Keller}}, \bibinfo {editor}
		{\bibfnamefont {J.~C.}\ \bibnamefont {Hensel}}, \ and\ \bibinfo {editor}
		{\bibfnamefont {F.}~\bibnamefont {Stern}}}\ (\bibinfo  {publisher} {USAEC
		Division of Technical Information},\ \bibinfo {address} {Washington D.C.},\
	\bibinfo {year} {1970})\ p.\ \bibinfo {pages} {271}\BibitemShut {NoStop}%
	\bibitem [{\citenamefont {Ramdas}\ and\ \citenamefont
		{Rodriguez}(1981)}]{Ramdas81}%
	\BibitemOpen
	\bibfield  {author} {\bibinfo {author} {\bibfnamefont {A.~K.}\ \bibnamefont
			{Ramdas}}\ and\ \bibinfo {author} {\bibfnamefont {S.}~\bibnamefont
			{Rodriguez}},\ }\href {http://stacks.iop.org/0034-4885/44/i=12/a=002}
	{\bibfield  {journal} {\bibinfo  {journal} {Reports on Progress in Physics}\
		}\textbf {\bibinfo {volume} {44}},\ \bibinfo {pages} {1297} (\bibinfo {year}
		{1981})}\BibitemShut {NoStop}%
	\bibitem [{\citenamefont {Zhukavin}\ \emph {et~al.}(2011)\citenamefont
		{Zhukavin}, \citenamefont {Kovalevsky}, \citenamefont {Tsyplenkov},
		\citenamefont {Shastin}, \citenamefont {Pavlov}, \citenamefont {H{\"u}bers},
		\citenamefont {Riemann}, \citenamefont {Abrosimov},\ and\ \citenamefont
		{Ramdas}}]{Zhukavin11}%
	\BibitemOpen
	\bibfield  {author} {\bibinfo {author} {\bibfnamefont {R.~K.}\ \bibnamefont
			{Zhukavin}}, \bibinfo {author} {\bibfnamefont {K.~A.}\ \bibnamefont
			{Kovalevsky}}, \bibinfo {author} {\bibfnamefont {V.~V.}\ \bibnamefont
			{Tsyplenkov}}, \bibinfo {author} {\bibfnamefont {V.~N.}\ \bibnamefont
			{Shastin}}, \bibinfo {author} {\bibfnamefont {S.~G.}\ \bibnamefont {Pavlov}},
		\bibinfo {author} {\bibfnamefont {H.-W.}\ \bibnamefont {H{\"u}bers}},
		\bibinfo {author} {\bibfnamefont {H.}~\bibnamefont {Riemann}}, \bibinfo
		{author} {\bibfnamefont {N.~V.}\ \bibnamefont {Abrosimov}}, \ and\ \bibinfo
		{author} {\bibfnamefont {A.~K.}\ \bibnamefont {Ramdas}},\ }\href@noop {}
	{\bibfield  {journal} {\bibinfo  {journal} {Applied Physics Letters}\
		}\textbf {\bibinfo {volume} {99}},\ \bibinfo {pages} {171108} (\bibinfo
		{year} {2011})}\BibitemShut {NoStop}%
	\bibitem [{\citenamefont {Wilson}\ and\ \citenamefont
		{Feher}(1961{\natexlab{a}})}]{Feher61}%
	\BibitemOpen
	\bibfield  {author} {\bibinfo {author} {\bibfnamefont {D.~K.}\ \bibnamefont
			{Wilson}}\ and\ \bibinfo {author} {\bibfnamefont {G.}~\bibnamefont {Feher}},\
	}\href {\doibase 10.1103/PhysRev.124.1068} {\bibfield  {journal} {\bibinfo
			{journal} {Physical Review}\ }\textbf {\bibinfo {volume} {124}},\ \bibinfo
		{pages} {1068} (\bibinfo {year} {1961}{\natexlab{a}})}\BibitemShut {NoStop}%
	\bibitem [{\citenamefont {Pines}\ \emph {et~al.}(1957)\citenamefont {Pines},
		\citenamefont {Bardeen},\ and\ \citenamefont {Slichter}}]{PBS57}%
	\BibitemOpen
	\bibfield  {author} {\bibinfo {author} {\bibfnamefont {D.}~\bibnamefont
			{Pines}}, \bibinfo {author} {\bibfnamefont {J.}~\bibnamefont {Bardeen}}, \
		and\ \bibinfo {author} {\bibfnamefont {C.~P.}\ \bibnamefont {Slichter}},\
	}\href {\doibase 10.1103/PhysRev.106.489} {\bibfield  {journal} {\bibinfo
			{journal} {Phys. Rev.}\ }\textbf {\bibinfo {volume} {106}},\ \bibinfo {pages}
		{489} (\bibinfo {year} {1957})}\BibitemShut {NoStop}%
	\bibitem [{\citenamefont {Lipari}\ \emph {et~al.}(1980)\citenamefont {Lipari},
		\citenamefont {Baldereschi},\ and\ \citenamefont {Thewalt}}]{Lipari80}%
	\BibitemOpen
	\bibfield  {author} {\bibinfo {author} {\bibfnamefont {N.}~\bibnamefont
			{Lipari}}, \bibinfo {author} {\bibfnamefont {A.}~\bibnamefont {Baldereschi}},
		\ and\ \bibinfo {author} {\bibfnamefont {M.}~\bibnamefont {Thewalt}},\ }\href
	{\doibase 10.1016/0038-1098(80)91152-7} {\bibfield  {journal} {\bibinfo
			{journal} {Solid State Communications}\ }\textbf {\bibinfo {volume} {33}},\
		\bibinfo {pages} {277} (\bibinfo {year} {1980})}\BibitemShut {NoStop}%
	\bibitem [{\citenamefont {Diarra}\ \emph {et~al.}(2007)\citenamefont {Diarra},
		\citenamefont {Niquet}, \citenamefont {Delerue},\ and\ \citenamefont
		{Allan}}]{Diarra07}%
	\BibitemOpen
	\bibfield  {author} {\bibinfo {author} {\bibfnamefont {M.}~\bibnamefont
			{Diarra}}, \bibinfo {author} {\bibfnamefont {Y.-M.}\ \bibnamefont {Niquet}},
		\bibinfo {author} {\bibfnamefont {C.}~\bibnamefont {Delerue}}, \ and\
		\bibinfo {author} {\bibfnamefont {G.}~\bibnamefont {Allan}},\ }\href
	{\doibase 10.1103/PhysRevB.75.045301} {\bibfield  {journal} {\bibinfo
			{journal} {Physical Review B}\ }\textbf {\bibinfo {volume} {75}},\ \bibinfo
		{pages} {045301} (\bibinfo {year} {2007})}\BibitemShut {NoStop}%
	\bibitem [{\citenamefont {Holland}\ and\ \citenamefont
		{Paul}(1962)}]{HollandPaul62}%
	\BibitemOpen
	\bibfield  {author} {\bibinfo {author} {\bibfnamefont {M.~G.}\ \bibnamefont
			{Holland}}\ and\ \bibinfo {author} {\bibfnamefont {W.}~\bibnamefont {Paul}},\
	}\href {\doibase 10.1103/physrev.128.30} {\bibfield  {journal} {\bibinfo
			{journal} {Physical Review}\ }\textbf {\bibinfo {volume} {128}},\ \bibinfo
		{pages} {30-38} (\bibinfo {year} {1962})}\BibitemShut {NoStop}%
	\bibitem [{\citenamefont {Samara}\ and\ \citenamefont
		{Barnes}(1987)}]{SamaraBarnes87}%
	\BibitemOpen
	\bibfield  {author} {\bibinfo {author} {\bibfnamefont {G.~A.}\ \bibnamefont
			{Samara}}\ and\ \bibinfo {author} {\bibfnamefont {C.~E.}\ \bibnamefont
			{Barnes}},\ }\href {\doibase 10.1103/physrevb.35.7575} {\bibfield  {journal}
		{\bibinfo  {journal} {Physical Review B}\ }\textbf {\bibinfo {volume} {35}},\
		\bibinfo {pages} {7575} (\bibinfo {year} {1987})}\BibitemShut {NoStop}%
	\bibitem [{\citenamefont {Perdew}\ \emph {et~al.}(1996)\citenamefont {Perdew},
		\citenamefont {Burke},\ and\ \citenamefont {Ernzerhof}}]{Perdew96}%
	\BibitemOpen
	\bibfield  {author} {\bibinfo {author} {\bibfnamefont {J.~P.}\ \bibnamefont
			{Perdew}}, \bibinfo {author} {\bibfnamefont {K.}~\bibnamefont {Burke}}, \
		and\ \bibinfo {author} {\bibfnamefont {M.}~\bibnamefont {Ernzerhof}},\ }\href
	{\doibase 10.1103/PhysRevLett.77.3865} {\bibfield  {journal} {\bibinfo
			{journal} {Phys. Rev. Lett.}\ }\textbf {\bibinfo {volume} {77}},\ \bibinfo
		{pages} {3865} (\bibinfo {year} {1996})}\BibitemShut {NoStop}%
	\bibitem [{\citenamefont {Bl\"ochl}(1994)}]{Blochl94}%
	\BibitemOpen
	\bibfield  {author} {\bibinfo {author} {\bibfnamefont {P.~E.}\ \bibnamefont
			{Bl\"ochl}},\ }\href {\doibase 10.1103/PhysRevB.50.17953} {\bibfield
		{journal} {\bibinfo  {journal} {Phys. Rev. B}\ }\textbf {\bibinfo {volume}
			{50}},\ \bibinfo {pages} {17953} (\bibinfo {year} {1994})}\BibitemShut
	{NoStop}%
	\bibitem [{\citenamefont {Kresse}\ and\ \citenamefont
		{Furthm\"uller}(1996)}]{Kresse96}%
	\BibitemOpen
	\bibfield  {author} {\bibinfo {author} {\bibfnamefont {G.}~\bibnamefont
			{Kresse}}\ and\ \bibinfo {author} {\bibfnamefont {J.}~\bibnamefont
			{Furthm\"uller}},\ }\href {\doibase 10.1103/PhysRevB.54.11169} {\bibfield
		{journal} {\bibinfo  {journal} {Phys. Rev. B}\ }\textbf {\bibinfo {volume}
			{54}},\ \bibinfo {pages} {11169} (\bibinfo {year} {1996})}\BibitemShut
	{NoStop}%
	\bibitem [{\citenamefont {Bl\"ochl}(2000)}]{Blochl00}%
	\BibitemOpen
	\bibfield  {author} {\bibinfo {author} {\bibfnamefont {P.~E.}\ \bibnamefont
			{Bl\"ochl}},\ }\href {\doibase 10.1103/PhysRevB.62.6158} {\bibfield
		{journal} {\bibinfo  {journal} {Phys. Rev. B}\ }\textbf {\bibinfo {volume}
			{62}},\ \bibinfo {pages} {6158} (\bibinfo {year} {2000})}\BibitemShut
	{NoStop}%
	\bibitem [{\citenamefont {Niquet}\ \emph {et~al.}(2010)\citenamefont {Niquet},
		\citenamefont {Genovese}, \citenamefont {Delerue},\ and\ \citenamefont
		{Deutsch}}]{Niquet10}%
	\BibitemOpen
	\bibfield  {author} {\bibinfo {author} {\bibfnamefont {Y.~M.}\ \bibnamefont
			{Niquet}}, \bibinfo {author} {\bibfnamefont {L.}~\bibnamefont {Genovese}},
		\bibinfo {author} {\bibfnamefont {C.}~\bibnamefont {Delerue}}, \ and\
		\bibinfo {author} {\bibfnamefont {T.}~\bibnamefont {Deutsch}},\ }\href
	{\doibase 10.1103/PhysRevB.81.161301} {\bibfield  {journal} {\bibinfo
			{journal} {Physical Review B}\ }\textbf {\bibinfo {volume} {81}},\ \bibinfo
		{pages} {161301} (\bibinfo {year} {2010})}\BibitemShut {NoStop}%
	\bibitem [{\citenamefont {Yu}\ and\ \citenamefont
		{Cardona}(1999)}]{YuCardona99}%
	\BibitemOpen
	\bibfield  {author} {\bibinfo {author} {\bibfnamefont {P.~Y.}\ \bibnamefont
			{Yu}}\ and\ \bibinfo {author} {\bibfnamefont {M.}~\bibnamefont {Cardona}},\
	}\href@noop {} {\emph {\bibinfo {title} {Fundamentals of semiconductors:
				physics and materials properties}}}\ (\bibinfo  {publisher} {Springer},\
	\bibinfo {year} {1999})\BibitemShut {NoStop}%
	\bibitem [{\citenamefont {Wilson}\ and\ \citenamefont
		{Feher}(1961{\natexlab{b}})}]{WilsonFeher61}%
	\BibitemOpen
	\bibfield  {author} {\bibinfo {author} {\bibfnamefont {D.~K.}\ \bibnamefont
			{Wilson}}\ and\ \bibinfo {author} {\bibfnamefont {G.}~\bibnamefont {Feher}},\
	}\href {\doibase 10.1103/physrev.124.1068} {\bibfield  {journal} {\bibinfo
			{journal} {Physical Review}\ }\textbf {\bibinfo {volume} {124}},\ \bibinfo
		{pages} {1068} (\bibinfo {year} {1961}{\natexlab{b}})}\BibitemShut {NoStop}%
	\bibitem [{YMN()}]{YMNshearnote}%
	\BibitemOpen
	\href@noop {} {}\bibinfo {note} {The form of Eq. (9) of Ref.
		\cite{WilsonFeher61} suggests that $\epsilon_{xy}$ in Ref.
		\cite{WilsonFeher61} is the ``engineering'' shear strain
		$\epsilon_{xy}=2\varepsilon_{xy}$, hence the factor 2 difference between Eq.
		(\ref{eqHshear}) and Eq. (8) of Ref. \cite{WilsonFeher61}}\BibitemShut
	{NoStop}%
	\bibitem [{\citenamefont {Rahman}\ \emph {et~al.}(2009)\citenamefont {Rahman},
		\citenamefont {Park}, \citenamefont {Boykin}, \citenamefont {Klimeck},
		\citenamefont {Rogge},\ and\ \citenamefont {Hollenberg}}]{Rahman09}%
	\BibitemOpen
	\bibfield  {author} {\bibinfo {author} {\bibfnamefont {R.}~\bibnamefont
			{Rahman}}, \bibinfo {author} {\bibfnamefont {S.~H.}\ \bibnamefont {Park}},
		\bibinfo {author} {\bibfnamefont {T.~B.}\ \bibnamefont {Boykin}}, \bibinfo
		{author} {\bibfnamefont {G.}~\bibnamefont {Klimeck}}, \bibinfo {author}
		{\bibfnamefont {S.}~\bibnamefont {Rogge}}, \ and\ \bibinfo {author}
		{\bibfnamefont {L.~C.~L.}\ \bibnamefont {Hollenberg}},\ }\href {\doibase
		10.1103/physrevb.80.155301} {\bibfield  {journal} {\bibinfo  {journal}
			{Physical Review B}\ }\textbf {\bibinfo {volume} {80}},\ \bibinfo {pages}
		{155301} (\bibinfo {year} {2009})}\BibitemShut {NoStop}%
\end{thebibliography}

\end{document}